\documentclass[twocolumn]{aastex62}

\usepackage{ulem}
\usepackage{appendix}
\usepackage{amsmath}
\usepackage{threeparttable}

\newcommand{\mb}{\boldsymbol}

\shorttitle{Dust Settling and Clumping}
\shortauthors{Xu \& Bai}

\begin{document}

\title{Dust Settling and Clumping in MRI Turbulent Outer Protoplanetary Disks}


\correspondingauthor{Ziyan Xu, Xue-Ning Bai}
\email{ziyanx@pku.edu.cn, xbai@tsinghua.edu.cn}

\author[0000-0002-2986-8466]{Ziyan Xu}
\affiliation{Kavli Institute for Astronomy and Astrophysics, Peking University, 100871 Beijing, China}
\affiliation{Department of Astronomy, Peking University, 100871 Beijing, China}

\author[0000-0001-6906-9549]{Xue-Ning Bai}
\affiliation{Institute for Advanced Study, Tsinghua University, 100084 Beijing, China}
\affiliation{Department of Astronomy, Tsinghua University, 100084 Beijing, China}

\begin{abstract}
Planetesimal formation is a crucial yet poorly understood process in planet formation. It is widely believed that planetesimal formation is the outcome of dust clumping by the streaming instability (SI).
However, recent analytical and numerical studies have shown that the SI can be damped or suppressed by external turbulence, and at least the outer regions of protoplanetary disks are likely weakly turbulent due to magneto-rotational instability (MRI).
We conduct high-resolution local shearing-box simulations of hybrid particle-gas magnetohydrodynamics (MHD), incorporating ambipolar diffusion as the dominant non-ideal MHD effect, applicable to outer disk regions.
We first show that dust backreaction enhances dust settling towards the midplane by reducing turbulence correlation time.
Under modest level of MRI turbulence, we find that dust clumping is in fact easier than the conventional SI case, in the sense that the threshold of solid abundance for clumping is lower.
The key to dust clumping includes dust backreaction and the presence of local pressure maxima, which in our work is formed by the MRI zonal flows overcoming background pressure gradient.
Overall, our results support planetesimal formation in the MRI-turbulent outer protoplanetary disks, especially in ring-like substructures.
\end{abstract}


\section{INTRODUCTION}

The process of planet formation encompasses a growth of more than 13 orders of magnitude in size. At the lower end, micron-sized dust particles stick together and coagulate into mm to cm sized solids. At the higher end, km-sized planetesimals grow into planetary cores, and eventually into terrestrial or giant planets of up to $10^5$ km in size. The intermediate stage involves the formation of km-sized planetesimals out of from mm-cm sized particles, and it is perhaps the least understood problem in planet formation (e.g., see review by \citet{chiang10}).

\subsection{Overview of planetesimal formation theory}

The obstacle in planetesimal formation arises primarily because of the existence of growth barriers. First, direct growth of dust beyond $\sim$mm-cm size has been found to be difficult, and encounters fragmentation and bouncing barriers \citep[e.g.,][]{blum08,guttler10,zsom10}. While various scenarios have been proposed to mitigate the problem \citep[e.g.,][]{windmark12,okuzumi12,kataoka13}, they generally rely on not-well-known dust surface properties, and more recent models can yield very different conclusions \citep[e.g.,][]{okuzumi19}. Second, even dust can grow larger, they suffer from the so-called ``meter-size barrier" \citep{weidenschilling77a}, or more generally the drift barrier \citep[e.g.,][]{birnstiel12}, so that dust must grow sufficiently quickly to bypass this barrier before they are lost through radial drift.

It is thus generally believed that planetesimal formation must proceed through some collective effect, where the responsible physical mechanism first gathers dust particles into dense clumps, which then collapse under their own self-gravity. Under this framework, the key to understanding planetesimal formation is to identify and characterize mechanisms that can strongly concentrate particles, typically of mm-cm in size as the end product of grain growth.

Particles with a broad range of sizes interact with gaseous protoplanetary disks (PPDs) via gas drag, characterized by the stopping time $t_{\rm stop}$. It is conventional to cast the stopping time in dimensionless form as $\tau_s\equiv\Omega t_{\rm stop}$, where $\Omega$ is the local orbital frequency of the disk. Particles are considered strongly coupled to the gas for $\tau_s\ll1$, and the coupling is marginal for $\tau_s\sim1$. Such interactions lead to the aforementioned radial drift, where an outward radial gas pressure gradient typically present in PPDs is felt by the gas but not dust. The gas thus rotates at sub-Keplerian velocity, while particles feel a head wind and hence drift radially inward. More generically, particles always drift towards higher pressure, and the effect is the most prominent for marginally coupled particles with $\tau_s\sim1$. Particles with mm-cm sizes
are typically marginally coupled with $\tau_s\sim0.1-1$ in the outer PPDs (a few tens of AU), and more strongly coupled in the inner disks ($\sim10^{-3}$ in the inner few AU). Therefore, mutual interaction between dust and gas is the key piece of physics in any theory of planetesimal formation.

Multiple planetesimal formation mechanisms have been proposed. Early studies focused on gravitational instability (GI) of a thin dust layer \citep[][]{safronov69,goldreich73}. However, it requires extreme level of settling, and the dust layer itself can be subject to the Kelvin-Helmholtz instability \citep[KHI,][]{weidenschilling80}. Triggering the GI requires unrealistically high dust-to-gas ratio \citep{sekiya98,youdin02}, although it can be mitigated to less extreme conditions \citep{chiang08,barranco09,lee10a,lee10b}. For strongly-coupled dust, concentration may be achieved through secular GI \citep[e.g.,][]{youdin11,takahashi14,tominaga18} and two-component viscous GI \citep{tominaga19}, resulting from gas drag and self-gravity over secular timescales, with the latter further involving the action of gas viscosity. They generally require very low level of external turbulence (viscosity) to operate. On the other hand, turbulence contains eddies of different sizes which themselves may selectively concentrate particles whose stopping time is comparable to eddy turnover time \citep[e.g.,][]{cuzzi01,pan11,hopkins16,hartlep20}. In addition, specific hydrodynamic instabilities can also result in strong vortices that act as dust traps \citep[e.g.,][]{raettig21}.

Over the past decade, perhaps the most popular and promising scenario for planetesimal formation is the streaming instability \citep[SI,][]{goodman00,youdin05}. The key ingredient behind the SI is the incorporation of backreaction from dust to gas, leading to a two-fluid type instability. It arises at the cost of the free energy from relative drift between dust and gas (which ultimately arises from the pressure gradient in the disk) rather than orbital shear. It is commonly interpreted as dust pile up causing traffic jams \citep{youdinj07,johansen07}, more recent investigations have revealed the richness of the instability in its linear behaviors \citep[][]{jacquet11,lin17,jaupart20,pan20a,pan20b,lin21}, and can be considered as a special case of the resonant drag instability \citep[][]{squire18,squire20}.

Non-linear simulations particle-gas dynamics in (vertically-stratified) PPDs have revealed that upon incorporating dust backreaction, the SI leads to the formation of dense dust clumps with sufficiently high density to drive gravitational collapse \citep[][]{johansen09b,johansen14,simonj16,schafer17,abod19}. The strength of particle concentration depends on physical conditions in PPDs, particularly particle stopping time, the solid abundance, and the background gas pressure gradient in the disk \citep[][]{bai10b,bai10c,carrera15,yang17}.
With these simulation results, the SI scenario has been widely applied to model planetesimal formation in global disk context  \citep[e.g.][]{drazkowska14,schoonenberg17}.

\subsection[]{The issue with disk turbulence}\label{ssec:turb}

The aforementioned studies of the SI typically consider dust dynamics without external turbulence. However, recent studies have shown that the linear growth of the SI can be damped or fully suppressed by even moderate level (Shakura-Sunyaev $\alpha\gtrsim10^{-4}$, \citealp{shakura73}) of disk viscosity \citep[][]{umurhan20,chen20}, a result that has been qualitatively confirmed from non-linear SI simulations with external driven turbulence \citep[][]{gole20}. This is a serious problem that endangers the applicability of the SI to planetesimal formation.

Observational inference of disk turbulence have been inconclusive, where there is evidence for both strong and weak turbulence.
Direct measurements of turbulent line broadening have yielded either strong \citep[][]{flaherty20} or non-detectable \citep[][]{flaherty17,flaherty18,teague18} level of turbulence. Level of turbulence can also be inferred from measuring dust distributions. In particular, dust layer thickness is determined by a balance of dust settling from vertical gravity and turbulent diffusion. In the case of the HL Tau disk \citep{alma15}, the sharpness of dust rings indicates a geometrically thin dusty disk corresponding to weak turbulence with $\alpha\sim$ a few times $10^{-4}$ \citep[][]{pinte16}.
Recent disk observations have also revealed rich substructures dominated by dust rings \citep[e.g. ][]{andrews18,long18}, where the width of dust rings result from dust trapping balancing turbulent diffusion. Modeling and measuring dust and gas ring width have led to a conclusion of relatively strong turbulence with $\alpha/\tau_s\gtrsim10^{-2}$, at least in disk outer regions \citep[][]{dullemond18,rosotti20}. Despite some controversies,
the inferred turbulence levels
are typically weak, but still well above $\alpha\sim10^{-4}$, which would have suppressed the SI.

An important caveat behind the conundrum above is that turbulence may not be simply approximated as a viscosity, nor mimicked by artificial turbulence driving. Depending on the source of turbulence driving, it may share substantially different properties. For instance, it was found that the SI can coexist with the vertical shear instability \citep[VSI, ][]{nelson13} which, despite yielding a turbulence level of $\alpha\sim10^{-3}$, can constructively act with the SI to promote dust clumping \citep[][]{schafer20}.
Therefore, it is of crucial importance to study the interplay of the SI with more realistic turbulence.

\subsection[]{This work}

In this work, we focus on the outer regions of PPDs, which are accessible to spatially-resolved observations. The most widely invoked mechanism for generating turbulence is the
magnetorotational instability (MRI, \citet{balbus91}), whereas long as gas and magnetic fields are well coupled, or in the ideal magnetohydrodynamics (MHD) regime, it generates vigorous turbulence that efficiently transports angular momentum and drives disk accretion. However, PPDs are weakly ionized, making gas and magnetic fields poorly coupled, resulting in three non-ideal MHD effects, namely, Ohmic resistivity, the Hall effect, and ambipolar diffusion (AD) \citep[e.g.][]{wardle07,bai11a,armitage15,lesur20}. The three effects dominate in progressively lower density regions, where the outer disk is largely dominated by AD. Simulations have revealed that the MRI is damped by AD \citep{bai11}, but not fully suppressed given the expected level of ionization in the outer disk \citep[][]{simon13a,simon13,bai15},
with $\alpha$ on the order of $10^{-3}$. It is our goal to study the interplay of the SI with the MRI in this regime.

We note that there are several similar studies in the literature considering particle concentration in MRI turbulence. Majority of the studies considered the MRI in ideal MHD, with low resolution (typically $\lesssim16$ cells per scale height), well below typical resolutions needed to properly resolve the SI \citep[e.g.][]{balsara09,tilley10}. Exceptions include \citet{johansen07Nat,johansen11}, who conducted high-resolution local MRI simulations in ideal MHD
and found particle concentration in local pressure bumps (known as zonal flows, \citet{johansen09a}), leading to planetesimal formation. Such zonal flows can be long-lived and likely play a significant role in concentrating particles \citep[][]{dittrich13}.
\citet{yang18} conducted large-box (but relatively low resolution) simulations with Ohmic resistivity in the midplane region (and hence modest turbulent damping), and also reported particle clumping despite that particles do not strongly settle to the midplane region. However, we anticipate the properties of the MRI turbulence to be very different in outer PPDs in the presence of AD.

Here, we perform very high resolution 3D simulations of particle-gas mixture in the local shearing-box framework. Both hydrodynamic and non-ideal MHD with AD are considered. Our simulations have both reasonable box size to accommodate the MRI turbulence and sufficient resolution to reasonably resolve the SI, if present, which allow us to study the interplay between realistic MRI turbulence and dust dynamics in the outer PPDs. We are particularly interested in particle settling and diffusion in vertical direction, as well as particle clumping. By comparison between runs with and without magnetic field, as well as with and without dust backreaction, we aim to decipher the potential mechanisms that control dust diffusion, settling and concentration, and whether the SI still operates to form planetesimals.

This paper is organized as follows. In \S \ref{sec:method}, we describe basic formulations and simulation setup. In \S \ref{sec:overview}, we overview the simulation results. In \S \ref{sec:parsettling}, \S \ref{sec:parclumping}, and \S \ref{sec:radtrans}, we analyze the simulation results, focusing on particle vertical settling and clumping respectively. The results are further analyzed and discussed in \S \ref{sec:discussion} to address the physics behind particle settling behaviour and particle clumping, as well as implications in realistic disks. We summarize and discuss future prospects in \S \ref{sec:conclusions}.

\begin{figure}[t]
\begin{center}
\includegraphics[width=0.47\textwidth]{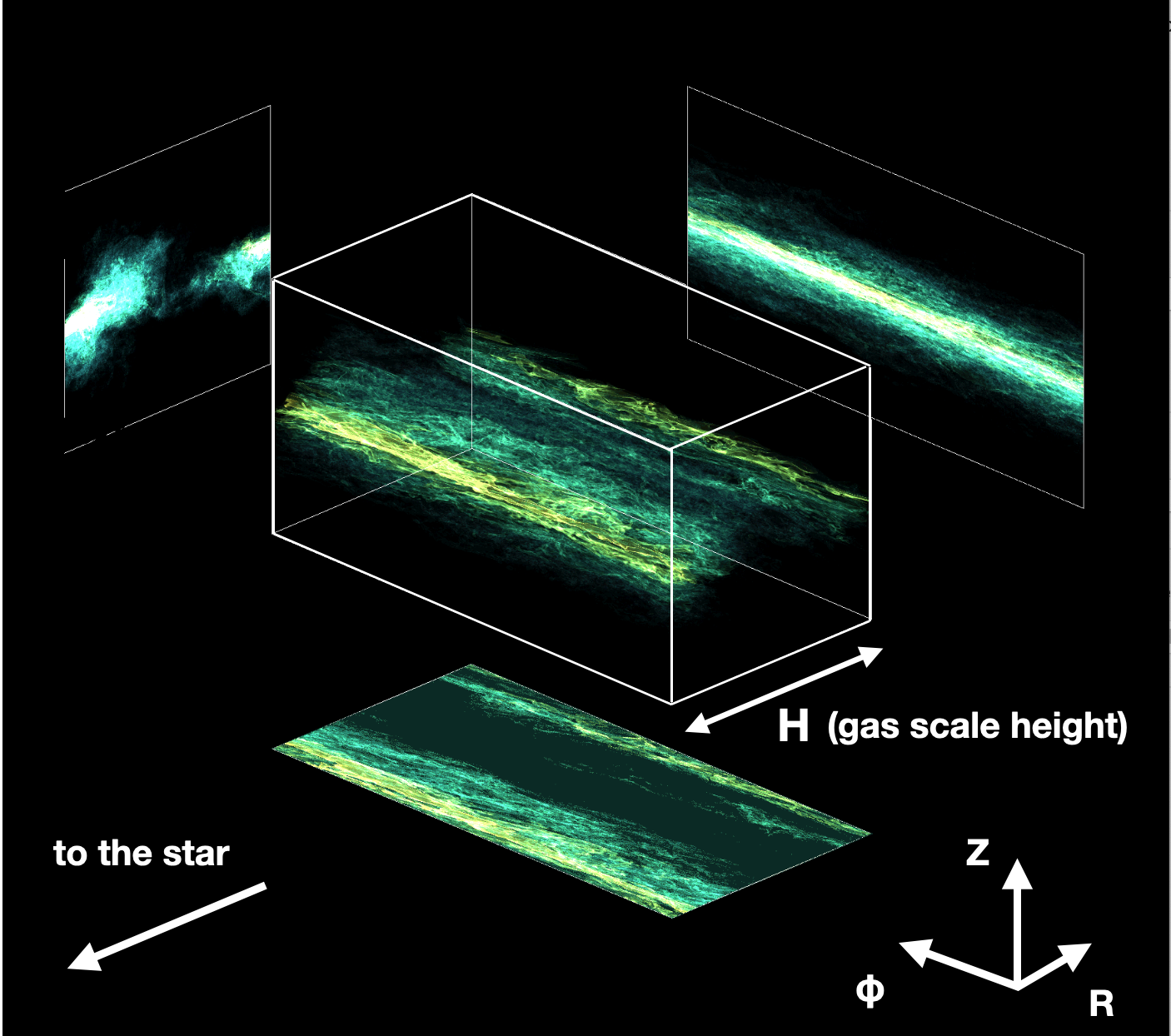}
\end{center}
\caption{A snapshot of 3D visualization of the dust spatial distribution in our non-ideal MHD shearing-box simulation (run ADZ2) at the time of $T=120\Omega_K^{-1}$ after inserting dust particles, with projections in three orthogonal directions. Dust density is mapped to brightness, where brighter region indicates higher dust density. \label{fig:simoverview}}
\end{figure}

\section{Method and Simulations} \label{sec:method}

\subsection{Formulation and simulation framework} \label{subsec:formulation}

We use the Athena MHD code \citep{stone08}, a higher-order Godunov code with constrained transport to preserve the divergence-free property for magnetic fields \citep{gardiner05,gardiner08} to perform 3D local MHD simulations in the shearing sheet approximation \citep{goldreich65}.
We use a local reference frame located at a fiducial radius corotating with Keplerian frequency $\Omega_K$. The dynamical equations are written in Cartesian coordinates, with $ \hat{\mb x}$, $\hat{\mb y}$ and $\hat{\mb z}$ denoting unit vectors pointing to the radial, azimuthal and vertical directions, and ${\mb\Omega}_K$ is along the $\hat{\mb z}$ direction. The gas density and velocity are denoted by $\rho_g$ and ${\mb u}$ separately in this non-inertial frame. The HLLD Riemann solver \citep{miyoshi05} with third order reconstruction, combined with the CTU integrator are used in all MHD simulations. Dust is implemented as Lagrangian particles coupled with gas via aerodynamic drag, and the backreaction from the particles to gas is included, as implemented in \citep{bai10}. The drag force between gas and dust particles is characterized by particle stopping time $t_{\rm stop}$. For a particle $i$ with velocity of ${\mb v_i}$, the drag force on the particle is $({\mb u}-{\mb v_i})/t_{\rm stop}$ per particle unit mass. With magnetic field denoted by ${\mb B}$, the MHD equations are written as:

\begin{equation}
  \label{eq.MHD1}
  \frac{\partial \rho_g}{\partial t} + \nabla \cdot (\rho_g {\mb u}) = 0,
\end{equation}
\begin{equation}
\begin{split}
\label{eq.MHD2}
&\frac{\partial \rho_g {\mb u}}{\partial t} + \nabla \cdot (\rho_g {\mb u}^T {\mb u}+ {\sf T}) \\
&=\rho_g [ 2 {\mb u} \times {\mb \Omega_K} + 3 \Omega_K^2 x {\hat{\mb x}}  + \epsilon \frac{{ \bar{\mb v}}-{\mb u}}{t_{\rm stop}}],
\end{split}
\end{equation}
\begin{equation}
\label{eq.MHD3}
\frac{\partial {\mb B}}{\partial t} = \nabla \times \left[ {\mb u} \times {\mb B} + \frac{({\mb J} \times  {\mb B}) \times  {\mb B}}{\gamma \rho_i \rho_g}\right],
\end{equation}
where ${\sf T}$ denotes the total stress tensor:
\begin{equation}
\label{eq.MHD4}
{\sf T} = (P + B^2/2){\sf I} - {\mb B}^T {\mb B},
\end{equation}
with ${\sf I}$ being the identity tensor and $P$ is the gas pressure. We assume an isothermal equation of state for the gas so that $P = \rho_g c_s^2$, where $c_s$ is the isothermal sound speed. The disk scale height is given by $H_g = c_s/\Omega_K$. The units for magnetic field is chosen such that magnetic permeability $\mu=1$, and hence magnetic pressure is given by $B^2/2$, and current density is given by ${\mb J}=\nabla\times{\mb B}$.

The first and second terms on the right side of equation (\ref{eq.MHD2}) represent Coriolis force and radial tidal gravity, separately. The third term on the right side of (\ref{eq.MHD2}) is the backreaction from dust particles to gas where $\epsilon$ and ${ \bar{\mb v}}$ are the local particle mass density and mean particle velocity, obtained by interpolation. The disk vertical gravity in the gas is ignored in our simulations and hence the gas is unstratified (but dust is stratified, see later). This is justified as we are mainly interested in regions within one gas scale height about the midplane.

In the induction Equation (\ref{eq.MHD3}), the last term corresponds to ambipolar diffusion (AD), where $\gamma$ is the coefficient of momentum exchange in ion-neutral collisions and $\rho_i$ denotes the ion density. AD is paraterized by Els{\"a}sser number:

\begin{equation}
\label{eq.Am}
Am = \frac{\gamma \rho_i}{\Omega_K},
\end{equation}
with $Am \to \infty$ corresponds to ideal MHD regime, where the magnetic field is frozen to the gas. The gas dynamics in the disk is significantly affected by AD with $Am \lesssim 10$ \citep{bai11}. \citet{bai11a,bai11b} found that $Am \sim 1$ is widely applicable at the outer PPDs ($\gtrsim$ 30 AU) where AD is the dominant non-ideal MHD effect.

Without stratification, initial gas density $\rho_0$ is uniform. We thus set $c_s = \Omega_K = H_g = \rho_0 = 1$ in code units.
Periodic boundary conditions are used in both vertical and azimuthal directions, while in the radial direction, we use shearing periodic boundary conditions \citep{hawley95}. An orbital advection scheme is used which improves the accuracy of the simulations \citep{stone10}, in which we subtract Keplerian shear ${\mb u}_K\equiv-(3/2)\Omega_Kx\hat{\mb y}$ and only evolve the deviation ${\mb u}'\equiv{\mb u}-{\mb u}_K$.

Fiducially, we impose a net vertical magnetic field $B_0$ in our MHD simulations. The strength of $B_0$ is characterized plasma $\beta$,

\begin{equation}
\label{eq.beta}
\beta_0 \equiv \frac{\rho_0 c_s^2}{B_0^2/2},
\end{equation}
corresponding to the ratio of gas pressure $\rho_0 c_s^2$ to the magnetic pressure of the net vertical field. The net vertical magnetic field is needed for the MRI turbulence to be sustained under strong AD in outer disk regions \citep{bai11,simon13a}, and $\beta_0 \sim 10^4$ is required for achieving accretion rates consistent with observations \citep{simon13,bai15}.

The equation of motion for particle $i$ can then be written as

\begin{equation}
\label{eq.dust1}
\frac{d{\mb v}_i}{dt} = -2\eta v_K \Omega_K {\hat{\mb x}} + 2 {\mb v}_i \times {\mb \Omega_K} + 3 \Omega_K ^2 x_i {\hat{\mb x}} - \Omega_K^2 z_i {\hat{\mb z}} -\frac{{ {\mb v}_i}-{\mb u}}{t_{\rm stop}}.
\end{equation}
In the above, the first term on the right side $-2\eta v_K \Omega_K {\hat{\mb x}}$ represents a force pointing inward in order to mimic the outward gas radial pressure gradient (see \citet{bai10}), which is equivalent to but more convenient than directly implementing the pressure gradient to the gas in Equation (\ref{eq.MHD2}). The parameter $\Delta v_K\equiv \eta v_K$ represents the deviation from Keplerian rotation due to radial pressure gradient, which is characterized by $\Pi\equiv\eta v_K/c_s$ \citep{bai10c}. In the absence of gas drag, $\Delta v_K$ gives the relative velocity between gas and dust, i.e., the headwind. With our treatment, gas would stay at Keplerian while dust would travel at super-Keplerian speeds, maintaining the relative drift between the two components.

\begin{table}[t]
    \centering
     \caption{List of simulation runs.}
    \label{tab:sumsetup}
    \begin{tabular}{lccccc}
   Run name  & $\Pi$ & $Z$ & $N_x \times N_y \times N_z$  & $\beta$ & $Am$\\
   \hline
   hydZ2 & 0.1 & 0.02 & 480 $\times$ 480 $\times$ 480 & $\infty$ & - \\
   hydZ4 & 0.1 & 0.04 & 480 $\times$ 480 $\times$ 480 & $\infty$ & - \\
   \hline
   ADZ0 & 0.1 & $10^{-5}$ & 256 $\times$ 256 $\times$ 256 & 12800 & 2 \\
   ADZ05 & 0.1 & 0.005 & 256 $\times$ 256 $\times$ 256 & 12800 & 2 \\
   ADZ1 & 0.1 & 0.01 & 256 $\times$ 256 $\times$ 256 & 12800 & 2 \\
   ADZ2 & 0.1 & 0.02 & 256 $\times$ 256 $\times$ 256 & 12800 & 2 \\
   ADZ4 & 0.1 & 0.04 & 256 $\times$ 256 $\times$ 256 & 12800 & 2 \\
      \hline
   ADZ1H & 0.1 & 0.01 & 384 $\times$ 384 $\times$ 384 & 12800 & 2 \\
   ADZ2H & 0.1 & 0.02 & 384 $\times$ 384 $\times$ 384 & 12800 & 2 \\
   \hline
   ADZ2P0 & 0 & 0.02  & 256 $\times$ 256 $\times$ 256 & 12800 & 2 \\
   ADZ2P005 & 0.05 & 0.02 & 256 $\times$ 256 $\times$ 256 & 12800 & 2 \\
   ADZ2P02 & 0.2 & 0.02 & 256 $\times$ 256 $\times$ 256 & 12800 & 2 \\

    \end{tabular}
    \begin{tablenotes}
    \item All simulations have a particle number of 1 per cell on average, with particle stopping time $\tau_s=0.1$.
    \item The simulation box size is $H \times 2H \times H$ for MHD runs, and $0.6H \times 1.2H \times 0.6H$ for hydrodynamic runs.
    \end{tablenotes}
\end{table}

The second and third terms on the right side of equation (\ref{eq.dust1}) again represent Coriolis force and radial tidal gravity, separately. Unlike for the gas, vertical gravity for dust particles is included in the simulation given by $-\Omega_K^2 z_i {\hat{\mb z}}$, thus allowing particles to settle towards the midplane. The last term represents the drag force on individual particles, where gas velocities are interpolated to individual particle positions.
It is convenient to use a dimensionless stopping time defined as $\tau_s \equiv \Omega_K t_{\rm stop}$, where particles with $\tau_s \ll 1$ are strongly coupled to the gas while particles with $\tau_s \gg 1$ are loosely coupled to the gas.

We use the second-order semi-implicit solver to integrate particles, with the triangular shaped cloud (TSC) scheme for interpolation (for dust integrator) and deposits (for dust backreaction), as described in \citet{bai10}.

\subsection{Simulation setup and parameters} \label{subsec:simsetup}

The list of our simulation runs is shown in Table \ref{tab:sumsetup}.
Fiducially, we perform non-ideal MHD simulations (labeled by ``AD") with particle feedback included, and an additional MHD run with sufficiently low solid abundance so that the dust feedback is negligible. We refer to the latter simulation as ``no feedback" (ADZ0), and rescale the particle density as appropriate when comparing with other runs with dust feedback. We also perform pure hydrodynamic simulations (labeled by ``hyd") with particle feedback (i.e. the standard SI simulations) for reference. In addition, we explore different levels of solid abundance, pressure gradient, turbulence levels and simulation resolutions. The simulation setup and parameters are described as follows.

As a first study, our simulations consider a single particle species with fixed stopping time $\tau_s=0.1$ for all our simulations. At outer disk regions, gas drag is in the Epstein regime with $t_{\rm stop} = \rho_s a / \rho_g c_s$ \citep{epstein24}, where $\rho_s$ and $a$ are the solid density and physical size of dust particles. Fixing $\tau_s$ is justified because gas density undergoes very little variation in unstratified MRI simulation box with AD \citep{bai11}. In standard disk models, $\tau_s=0.1$ corresponds to mm-cm size particles in the midplane regions of the outer disk \citep[e.g.,][]{chiang10}, which are also the most active in driving the SI \citep[e.g.,][]{bai10b}.

We consider a range of particle abundances, measured by the height-integrated dust-to-gas mass ratio $Z$, ranging from $Z=0.005$ to $Z=0.04$. We also consider a run with $Z=10^{-5}$ where there is essentially no particle backreaction. These runs are labeled by ``$Z$" followed by a number indicating the $Z$ values. Note that in interpreting $Z$, we assume the gas surface density is given by $\Sigma_g=\sqrt{2\pi}\rho_0H_g$ as if the gas is vertically stratified. The height-integrated dust mean surface density is thus $\Sigma_d=Z\Sigma_g$. This particle abundance is then evenly divided to individual particle ``mass density" in the simulations.

For gas radial pressure gradient, we choose $\Pi = 0.1$ by default, which is generally applicable at outer disk ($\gtrsim$ 30 AU) \footnote{For standard minimum-mass solar nebular \citep[MMSN,][]{weidenschilling77a} model with $\Sigma(R) =  1700 (R/{\rm AU})^{-3/2}\ {\rm g}\ {\rm cm}^{-2}$ and $T(R) = 280 (R/{\rm AU})^{-1/2}{\rm K}$, we have \citep[e.g.,][]{xu17}
\begin{equation}
\frac{\Delta v_K}{c_s} \approx 0.127 \left( \frac{R}{30{\rm AU}} \right)^{1/4}\ .
\end{equation}
The observationally inferred disk surface distribution in the main body of the disk is shallower ($\Sigma \sim R^{-1}$ instead of $R^{-3/2}$, e.g., \citet{andrews09,andrews10}, which lowers $\Delta v_K / c_s$ to $\sim 0.1$ at 30 AU.}.
We further consider simulations with different values of $\Pi=$ 0, 0.05, 0.1, 0.2 for comparison. These runs are labeled by ``$P$" followed by a number indicating the $\Pi$ values.

In our MRI simulations, we fix $Am = 2$ for our simulations. This makes turbulence slightly stronger than that with $Am=1$ so that we can better resolve the dust particle layer, while still maintaining a realistic turbulence level. We also fix the net vertical field with $\beta_0 = 1.28 \times 10^4$. This yields a most unstable MRI wavelength to be about $\sim0.12H$ (\citet{bai11}, see their Equation 22).

We choose simulation box size to be $H\times 2H \times H$ in $x$, $y$ and $z$ dimensions respectively, with a fiducial grid size of $256^3$. Thus, we can well resolve the MRI with $\sim30$ cells per most unstable wavelengths. Note that the resolution in the $y-$ axis is half of that in other dimensions due to the anisotropic nature of the MRI turbulence \citep[e.g., ][]{hawley11}. Such resolution is also sufficient to resolve the SI given $\Pi=0.1$, with $\sim 26$ cells in a characteristic length of $\Pi H$ for the SI. To our knowledge, this resolution is also higher (if not much higher) than all previous MRI simulations with particles (closest to ours is \citet{johansen11}, where their high-resolution run resolves the length of $\Pi H$ with $\sim 20$ cells). We also consider a higher resolution run with a grid size of $384^3$ for comparison.
For reference, we also conduct two pure hydrodynamic simulations, with smaller simulation box size of $0.6H \times 1.2H \times 0.6H$ and higher resolution with $480^3$ grid points. The main reason for using even higher resolution is to properly resolve the thinner dust layer in this case.

For MHD simulations, we first run simulations without particles for a time of $t_0 = 120 \Omega_K^{-1}$ with initially seeded small-amplitude random velocity perturbations in order for the MRI to grow and fully saturate into turbulence.
Particles are then inserted into the simulation box with random positions but with vertical distribution following a Gaussian profile $\propto \exp(-z^2/2H_{\rm p0}^2)$, with $H_{\rm p0} = 0.02 H_g$. The particle velocity is initialized according to Nakagawa-Sekiya-Hayashi (NSH) equilibrium \citep{nakagawa86}.
For pure hydrodynamic simulations, we use the same procedure except that particles are injected at the beginning.
For the remaining of this paper, we use the time $T$ after particle injection to denote simulation time.
In all simulations, there are on average one particle per cell. As particles settle towards the midplane, allowing us to build up sufficient particle statistics in the simulations to study the dynamics and clumping in the particle layer.

\begin{figure*}[t]
\begin{center}
\includegraphics[width=0.85\textwidth]{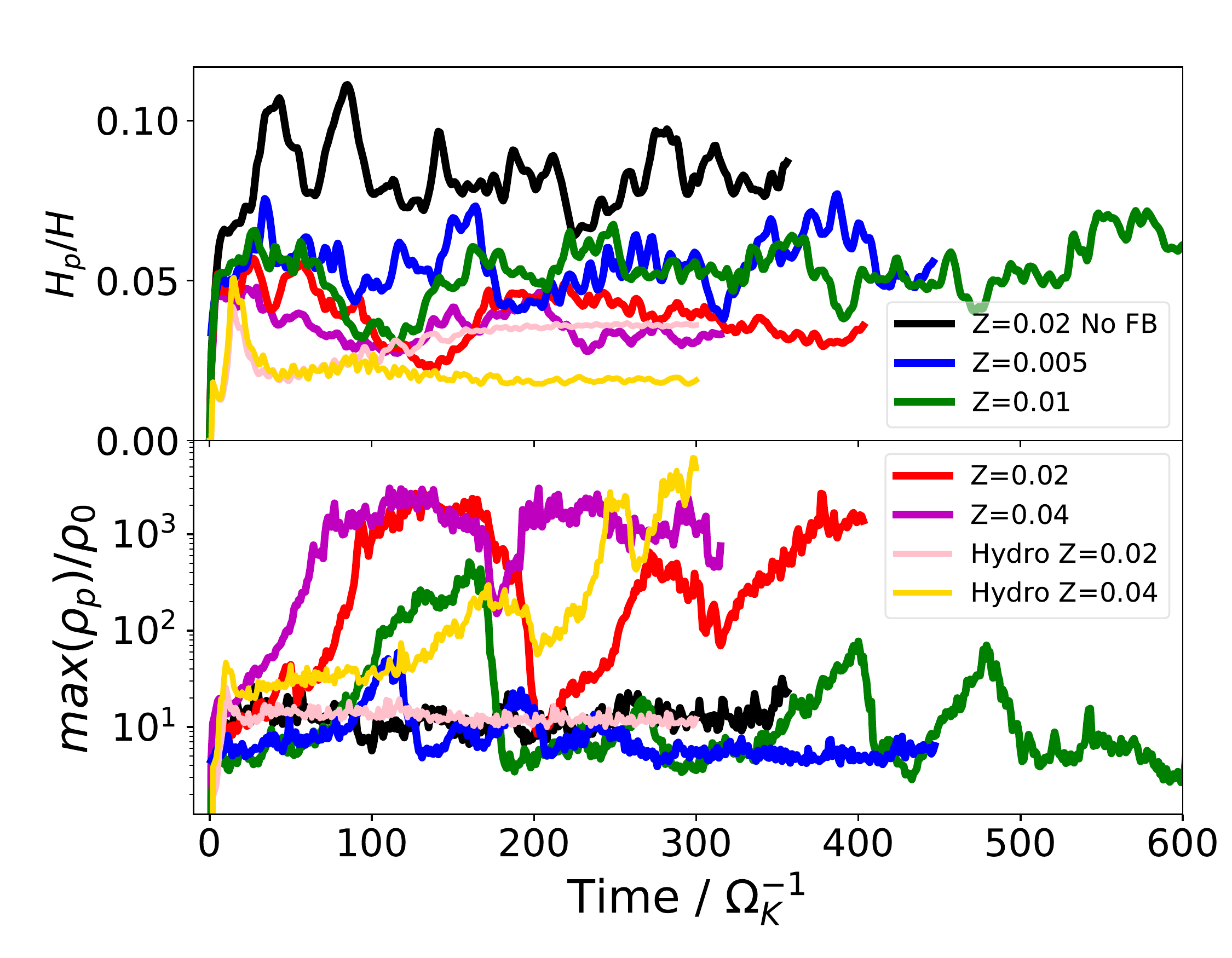}
\end{center}
\caption{\label{fig:hpdmax_hist} Time evolution of particle scale height (upper panel) and particle maximum density (lower panel) for simulation runs with $\Pi = 0.1$ and fiducial resolution, namely, runs ADZ0, ADZ05, ADZ1, ADZ2, ADZ4, hydZ2, and hydZ4. ``No FB'' stands for no feedback. The particle maximum density for run ADZ0 (shown with black lines) is scaled to $Z=0.02$. The horizontal axis represents time after inserting particles.
}
\end{figure*}

\section{Overview of simulations}\label{sec:overview}

Before presenting more detailed analysis and discussions of our simulation results, an overview of our simulations is shown in Figure \ref{fig:simoverview}, with a 3D visualization of the dust spatial distribution in our fiducial non-ideal MHD simulation (ADZ2) at the time of $T=120\Omega_K^{-1}$ after particle injection.
The figure highlights that dust settles to the midplane, maintaining a layer with scale height $H_p$ due to a balance between vertical gravity and turbulent diffusion. The dust layer is also radially segregated, which is due to the formation of zonal flows, as will be discussed later.

\paragraph{Dust Settling}
The upper panel of Figure \ref{fig:hpdmax_hist} shows the evolution of the normalized particle scale height $H_p/H$ over time, for all the runs with pressure gradient of $\Pi = 0.1$, where $H_p$ is measured using the rms value of the vertical coordinate for all the particles.

For the MRI turbulent run without particle feedback, the particle layer is puffed up from the initial setup and saturates within the time of $\sim 30 \Omega_K^{-1}$. The resulting $H_p\sim 0.08 H$ is comparable with that from the non-ideal MHD simulation with AD for $\tau_s=0.1$ in \citet{zhu15} and \citet{xu17}.
With particle feedback included in the simulations, the particle scale height in the steady state is lower than that without feedback, ranging from $H_p \sim 0.03 H$ (resolved by $\sim$16 cells) to $H_p \sim 0.06 H$.
This reduction of dust layer thickness indicates that particle feedback enhances dust settling in MRI turbulent disks.

For comparison, our pure hydrodynamic runs saturate at the time of $\sim 150 \Omega_K^{-1}$, with $H_p \sim 0.03 H$ and $H_p \sim 0.02 H$ for hydZ2 and hydZ4, separately.
The $H_p$ in the no-feedback (pure MRI) run is $\sim 3$ times higher than in the hydrodynamic (pure SI) runs,
while with feedback included, $H_p$ in AD case is only slightly/modestly higher than in the hydrodynamic case.
We will further discuss the vertical transport of dust particles and the role of dust feedback in dust settling in \S \ref{sec:parsettling}.

\paragraph{Dust Concentration}
In addition to the formation of dust filament seen in Figure \ref{fig:simoverview},
we further show in the lower panel of Figure \ref{fig:hpdmax_hist} the particle maximum density in the simulations as a function of time.

With dust feedback included, particle maximum density is significantly enhanced in our MHD simulations, especially when the solid abundance reaches $Z\gtrsim0.02$ (ADZ2 and ADZ4), with local particle density reaching up to 1000 times background gas density at the midplane, sufficient to trigger planetesimal formation if self-gravity is included. For comparison, particle clumping with similar particle maximum density is seen at $Z=0.04$ for the pure hydrodynamic case, but not for the $Z=0.02$ case. For AD runs with lower solid abundances (ADZ1 and ADZ05), particle concentration is much weaker and the maximum particle local density is rarely exceed a hundred times of the midplane gas density.

We will further analyze particle concentration in \S \ref{sec:parclumping}, and discuss the role of pressure variations induced by zonal flows on particle concentration.

\paragraph{Level of turbulence}
Both the dust layer thickness and the concentration of dust particles can be largely affected by the turbulence level in the disk.
The level of MRI turbulence in our simulation can be characterized by $\alpha$ parameter for disk angular momentum transport \citep{shakura73}, which is obtained by evaluating the time- and volume-averaged sum of the Maxwell stress and Reynolds stress, normalized by thermal pressure:

\begin{equation}
    \alpha \equiv \frac{-B_xB_y+\rho v_x v_y}{\rho c_s^2}.
\end{equation}

Our non-ideal MHD simulation without feedback (ADZ0) gives $\alpha=8.3\times 10^{-4}$, which is consistent with previous studies \citep{zhu15,xu17}. On the other hand, more direct characterization of level of turbulence arises from simply computing the rms velocities for each velocity components. We find from the ADZ0 run that $\langle v_x^2/c_s^2 \rangle = 1.2 \times 10^{-3}$, $\langle v_y^2/c_s^2 \rangle = 1.3 \times 10^{-3}$, and $\langle v_z^2/c_s^2 \rangle = 6.2 \times 10^{-4}$, comparable with the results from \citet{zhu15}. Simulations with dust feedback included generally give such rms velocities that are close to that from run ADZ0.

\section{Particle vertical transport} \label{sec:parsettling}

One important finding from our simulations is the reduction of particle scale height when considering particle feedback. More quantitatively, here we list the $H_p$ measured in our simulations, time-averaged from $T=60\Omega_K^{-1}$ to the end of each run, being 0.082$H$, 0.055$H$, 0.051$H$, 0.037$H$ and 0.034$H$ for runs ADZ0 (no feedback), ADZ05, ADZ1, ADZ2 and ADZ4, respectively. In this section, we diagnose in more detail about the cause of this reduction, from two separate perspectives, discussed in the following two subsections.

As a preliminary, we expect particle scale height $H_p$ to be related to both the turbulence properties and the particle stopping time $\tau_s$, and can be written as \citep{youdin07}

\begin{equation}  \label{eq:hp0}
  H_p = \sqrt{\frac{\langle v_{z}\rangle^2 \tau_e}{\Omega^2 \tau_s}} \cdot \sqrt{\frac{\tau_s+\tau_e}{\tau_s+\tau_e+\tau_s \tau_e^2}},
\end{equation}
where the dimensionless eddy time $\tau_e \equiv \Omega_K t_{\rm eddy}$ is defined for the convenience of characterizing the turbulence eddy time $t_{\rm eddy}$ (formally defined in \S \ref{subsec:teddy}).  The turbulence properties control particle scale height with both the gas vertical rms velocity $\langle v_{z}\rangle$ and the eddy time $t_{\rm eddy}$. The gas vertical diffusion coefficient can be written as

\begin{equation} \label{eq:dz}
  D_{g,z}=\langle v_{z}\rangle^2 t_{\rm eddy}.
\end{equation}

For dust particles with Stokes number $St \equiv \tau_s / \tau_e \ll 1$ and $\tau_s  \tau_e \ll 1$, the second term of Equation (\ref{eq:hp0}) is approximately unity, and Equation (\ref{eq:hp0}) is reduced to

\begin{equation}\label{eq:hp_dz}
  H_p \sim \sqrt{\frac{D_{g,z}}{\Omega \tau_s} } = \sqrt{\frac{\alpha_z}{\tau_s}} H,
\end{equation}
where $\alpha_z \equiv D_{g,z} /(c_s H)$ is a dimensionless parameter defined for characterizing vertical turbulent diffusion.

\subsection{Vertical profiles} \label{subsec:vertical}

\begin{figure}[t]
\begin{center}
\includegraphics[width=0.49\textwidth]{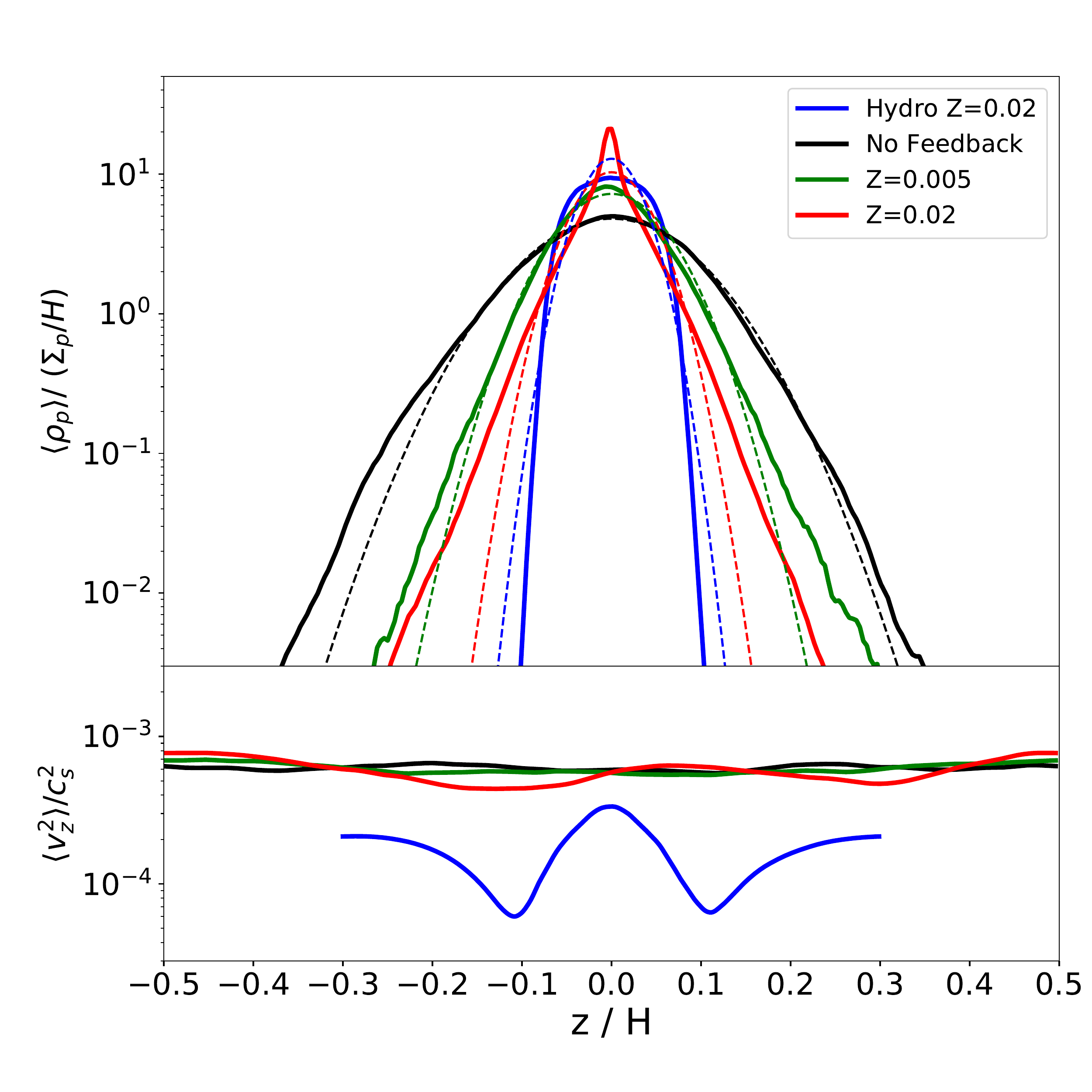}
\end{center}
\caption{\label{fig:zprofile} Vertical profiles of particle density (upper panel) and gas vertical rms velocity (lower panel) for runs hydZ2, ADZ0, ADZ05, and ADZ2. Solid lines are time and horizontally averaged vertical profiles in the saturate states of the runs. Dashed lines in the upper panel are the Gaussian profiles assuming time-averaged particle scale heights from each run.}
\end{figure}

If the dust vertical profile is Gaussian, it is uniquely characterized by $H_p$. However, this is not necessarily the case. In Figure \ref{fig:zprofile}, we show the vertical profiles of particle density (upper panel) and the gas vertical rms velocity (lower panel), in order to further study particle vertical transport.

For MHD simulation without particle feedback (ADZ0), the particle density profile generally agrees with the Gaussian profile, especially close to the disk midplane. For run ADZ05, Where particle feedback is included but no particle clumping is seen ($\max(\rho_p)/\rho_0$ always lower than 100), the dust density profile is overall Gaussian (except for more extended wings), but becomes narrower than that for run ADZ0.

Despite this reduction in particle scale height, the lower panel of Figure \ref{fig:zprofile} shows that the gas vertical rms velocity $\langle v^2_z \rangle$ in non-ideal MHD simulations with and without dust feedback is highly consistent throughout the disk, with $\langle v^2_z \rangle / c_s^2 \sim 0.0006$.
This indicates that dust feedback is not controlling the particle layer thickness in MRI turbulent disks by affecting the $\langle v^2_z \rangle$. Recalling Equation (\ref{eq:hp0}), we thus need to further examine effect of the turbulence eddy time (see next subsection).

For run ADZ2 (and similarly ADZ4 but not shown), there is further reduction in $H_p$, but more interestingly, the vertical density profile of particles is clearly non-Gaussian, as seen in Figure \ref{fig:hpdmax_hist}. A cusp is present at the midplane in the density vertical profile, and besides the cusp, the density profile is roughly exponential rather than Gaussian. Note that this density profile is measured over time $T=100-200\Omega_K^{-1}$ after particle injection, where strong dust clumping is present, and such deviations are likely the direct consequence of dust clumping.
However, dust clumping and additional settling does not react back to the gas strong enough to affect the vertical rms velocity of the gas, and we see that $\langle v^2_z \rangle$ remains vertically uniform throughout the disk.

The results obtained above are in contrast with our pure hydrodynamic simulations.
In run hydZ2, the vertical dust diffusion is generated by the SI turbulence itself. We see that the vertical rms velocity peaks at the midplane, which is natural as most particles that drive the SI reside in the midplane region, with $\langle v^2_z \rangle / c_s^2 \sim 0.0003$, slightly lower than that of the MHD runs. It then falls off away from the midplane \footnote{Away from the midplane, the increase of $\langle v^2_z \rangle$ is likely attributed to the numerical artifact from periodic vertical boundary conditions, as discussed in \citet{li18}, but as very few particles reside at larger vertical heights, its impact on particle dynamics is minimum.}. As a result, the particle vertical profile becomes more extended than the Gaussian profile at the midplane, but rapidly drops away from the midplane. This result is similar to those from previous SI simulations \citep[e.g.][]{bai10b}. Overall, comparison of our hydrodynamic and MRI runs shows that the presence of external turbulence has a strong effect on dust dynamics.

\subsection{The effect of $t_{\rm eddy}$} \label{subsec:teddy}

The fact that dust settles more strongly when turning on feedback, while $\langle v^2_z \rangle$ remains largely unchanged raises further question to the origin of the reduced particle vertical diffusion. In addition to the mean square gas vertical velocity $\langle v_z^2 \rangle$, particle diffusion in turbulent gas disk is also determined by the correlation time of turbulent fluctuations, defined as

\begin{equation}
t_{\rm eddy} \equiv \int_{0}^{\infty} R_{zz}(\tau)/R_{zz}(0)d\tau,
\end{equation}
where $R_{zz}(\tau) \equiv \langle v_z(\tau)v_z(0) \rangle$ is the auto correlation function for $v_z$ at a timescale of $\tau$. Although conventional ideal MHD simulations suggest $t_{\rm eddy} \sim \Omega_K^{-1}$ \citep{fromang06,carballido11}, \citet{zhu15} found that the correlation time in the vertical direction is increased in non-ideal MHD simulations with AD.

\begin{figure}[t]
\begin{center}
\includegraphics[width=0.47\textwidth]{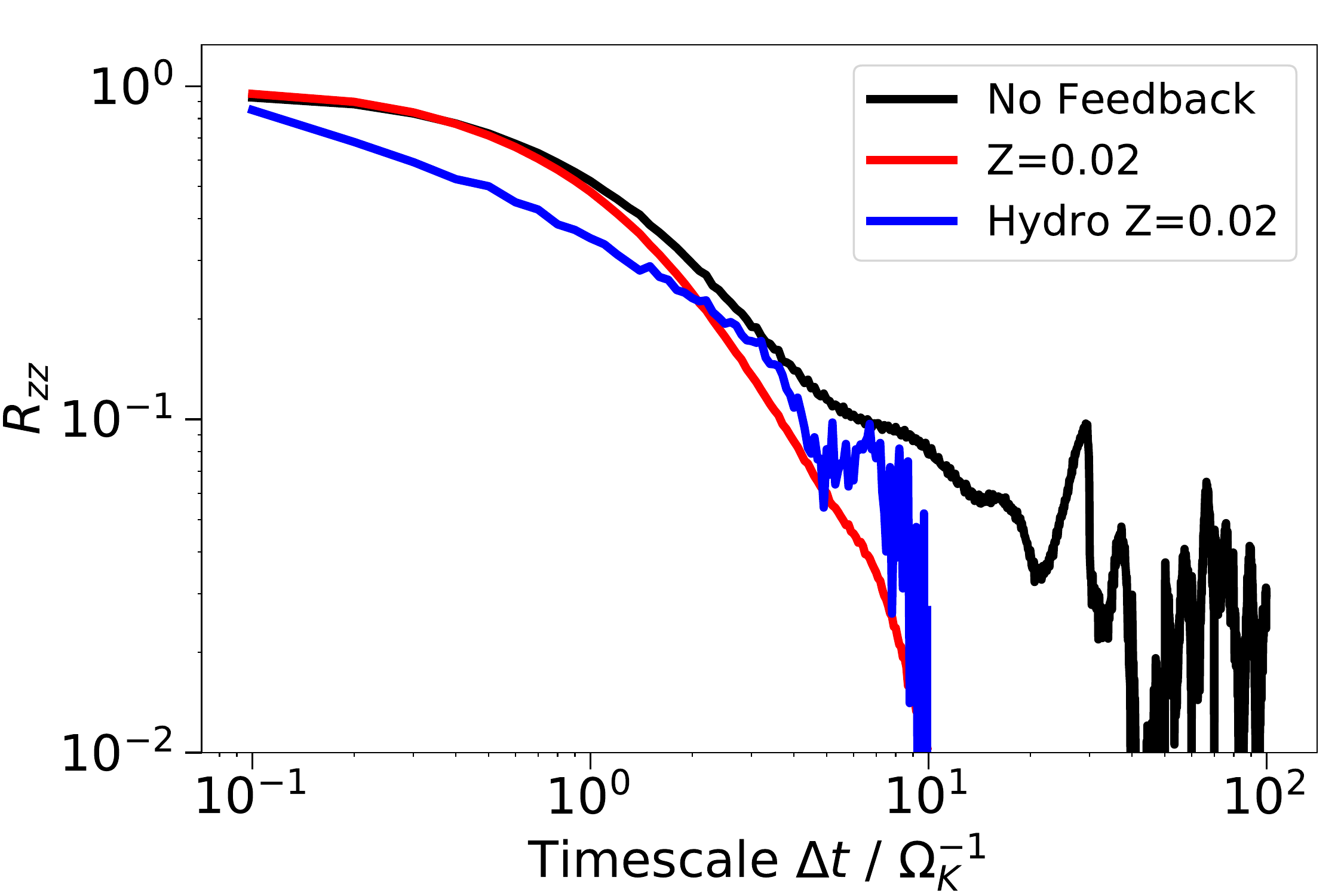}
\end{center}
\caption{\label{fig:rzz_time} Auto-correlation functions of gas vertical velocity fluctuations for runs ADZ0, ADZ2, and hydZ2 as a function of time scale. Without dust feedback, the eddy time for run ADZ0 is $t_{\rm eddy} \sim 5\Omega_{K}^{-1}$. With dust feedback in ADZ2, the eddy time is reduced and is comparable to the pure hydrodynamic case ($t_{\rm eddy} \sim 1.5\Omega_{K}^{-1}$).}
\end{figure}

In order to investigate the effect of particle feedback on the turbulent correlation time, we calculate the auto-correlation functions for $v_z$ in our simulations. We first output our simulation data with time intervals down to 0.1$\Omega_K^{-1}$, and shift the data along the azimuthal direction with a distance of 1.5$\Omega_K x t$ to correct for Keplerian shear.
The auto-correlation function $R_{zz}(\Delta t)$ is then derived by averaging $v_z(t)v_z(t+\Delta t)$ over time and space. The results are shown in Figure \ref{fig:rzz_time}.

For pure hydrodynamic simulations, the integration of the auto correlation function gives $t_{\rm eddy} \approx 1.5\Omega_K^{-1}$, which is consistent with the conventional expectation of $t_{\rm eddy} \sim \Omega_K^{-1}$. For non-ideal MHD simulation without particle feedback, the auto correlation function extends to longer timescales and the correlation time is increased to $t_{\rm eddy} \approx 5.0\Omega_K^{-1}$ due to the ambipolar diffusion, overall consistent with \citet{zhu15}. When particle is included in the non-ideal MHD simulation, we find that the auto correlation function drops rapidly at $\Delta t > 2 \Omega_K^{-1}$, and the resulting $t_{\rm eddy} \approx 1.5\Omega_K^{-1}$, close to the measured eddy time in the pure hydrodynamic simulation. With the measured $t_{\rm eddy}$ and $\langle v_z \rangle^2$, the vertical diffusion coefficients can be calculated with Equation (\ref{eq:dz}). Our calculation gives $D_{g,z} \sim 0.003 $ for ADZ0, and $D_{g,z} \sim 9 \times 10^{-4}$ for ADZ2. For pure hydrodynamic case, turbulent diffusion is inhomogeneous and is strongest at midplane with $\langle v^2_z \rangle / c_s^2 \sim 3 \times 10^{-4}$, corresponding to $D_{g,z} \sim 4.5 \times 10^{-4}$.
Following equation (\ref{eq:hp_dz}), our inferred $D_{g,z}$ would give $H_p\approx0.17H$ for ADZ0, and $H_p\approx0.095H$ for ADZ2. Both of these are higher than $H_p$ measured in the simulation by a factor of $\sim2$, which may reflect the complex nature of realistic turbulence, but the values are consistent with the fact that the particle layer thickness is reduced by $\sim$ half in ADZ2 compared to ADZ0.

We conclude that in MRI turbulent disks under AD, particle feedback likely enhances particle settling by reducing turbulence correlation time, instead of controlling the gas rms velocity.
We emphasize that our results are applicable only in the MRI turbulence with AD, where $t_{\rm eddy}$ is increased compared to ideal MHD case \citep{zhu15}.
We have also tested (but not presented) simulations with ideal MHD, and such reduction in eddy time is not observed when dust feedback is included.
This comparison further indicates that dust feedback reduces the turbulence correlation time by canceling out the enhancement of correlation time due to AD.

\section{Particle Concentration} \label{sec:parclumping}

\subsection{Particle clumping} \label{subsec:parclumping}

While our simulations do not include self-gravity, we expect that gravitational collapse can be triggered to form planetesimals when dust particles are sufficiently concentrated in high-density clumps. The standard criterion for gravitational collapse is that self-gravity of the dust clump must overcome the stellar tidal shear, i.e., dust density in the clump must exceed the Roche density:

\begin{equation} \label{eq:roche}
{\rho_R} = \frac{9 M_{\star}}{4\pi a^3},
\end{equation}
where $M_{\star}$ denotes the mass of the central star, and $a$ is the distance from the central star.
Assuming a standard MMSN disk, the gas density at the midplane is given by \citep{hayashi81}

\begin{equation} \label{eq:mmsnrho0}
\rho_0 = 1.4 \times 10^{-9} (a/AU)^{-2.75} {\rm g ~ cm}^{-3}.
\end{equation}
Combining Equations (\ref{eq:roche}) and (\ref{eq:mmsnrho0}), the corresponding local dust-to-gas ratio for a solar-mass star is

\begin{equation}
    \frac{\rho_R}{\rho_0} = 3.03\times10^{2} (a/AU)^{-0.25} {\rm g~cm}^{-3},
\end{equation}
At 1 AU, ${\rho_R}/{\rho_0}$ is approximately 300, and ${\rho_R}/{\rho_0}\sim 130$ at outer disk around 30 AU.

In the lower panel of Figure \ref{fig:hpdmax_hist}, the evolution of particle maximum density indicates strong clumping for solid abundances of $Z=0.02$ and $0.04$ in the presence of MRI turbulence and dust feedback, as ${\rm max}(\rho_p)$ reaches $10^3$ larger than the gas density at midplane, well exceeding the estimated Roche density. For lower solid abundances of Z=0.01 and Z=0.005, particle concentration is much weaker and the maximum solid density is generally lower than the Roche density.
Also note that in run ADZ2, the particle maximum density temporarily drops after $T \sim 200\Omega_K^{-1}$, indicating that the clump is disrupted.
This may be attributed to the fact that dust self-gravity is not included in our simulations, and that $Z=0.02$ is very close to the threshold solid abundance for dust clumping (see next subsection for more clues).
We will further discuss the survival of dust clumps and collapse criterion in \S \ref{subsec:clumpingcrit}.

It is useful to compare the results of particle clumping in our MHD simulations with those in pure hydrodynamic simulations. In our pure hydrodynamic simulation, particle clumping is seen in for $Z=0.04$, but not seen for $Z=0.02$. Surprisingly, this critical $Z$ for clumping is higher than that in our MHD simulations.
In general, dust abundance $Z$ above solar is needed to trigger clumping from the SI \citep{johansen09b,bai10b}.
Assuming a background gas pressure gradient of $\Pi = 0.05$, the clumping threshold for particles with $\tau_s\approx0.1$ is about $Z\gtrsim0.02$ (\citealp{carrera15}, \citealp{yang17}, but see \citealp{li21} who found a lower $Z$ threshold with improved numerics). However, our simulation adopts a higher pressure gradient of $\Pi=0.1$ that is likely more applicable to the outer disk, and our result of dust clumping threshold at $Z=0.04$ is generally consistent with the study of pressure gradient versus critical solid abundance \citep{bai10c,sekiya18}.

Overall, our simulations show that under more realistic MRI turbulence at the outer disk, particle clumping generally requires lower solid abundance (i.e., easier!) compared to conventional SI studies in pure hydrodynamic case, and the trend that clumping favors higher Z still holds in our MHD simulations.

\begin{figure*}[t]
\includegraphics[width=0.33\textwidth]{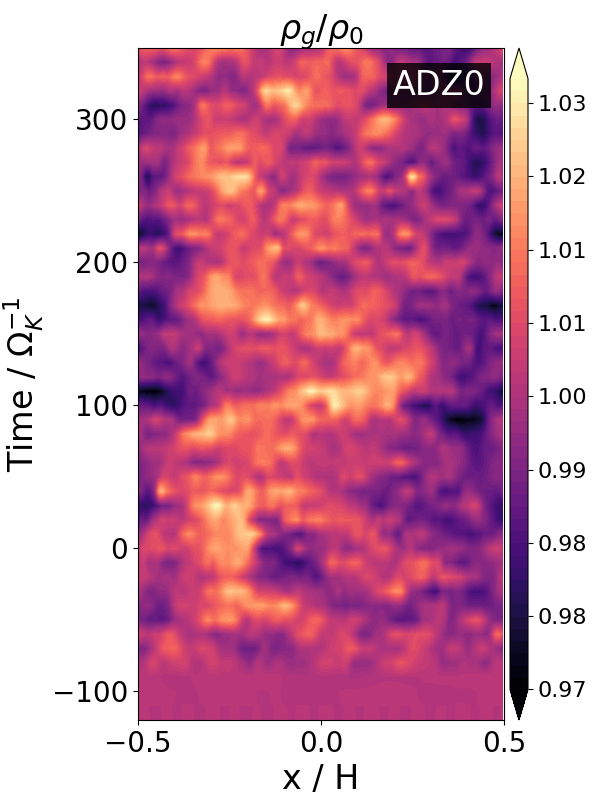}
\includegraphics[width=0.33\textwidth]{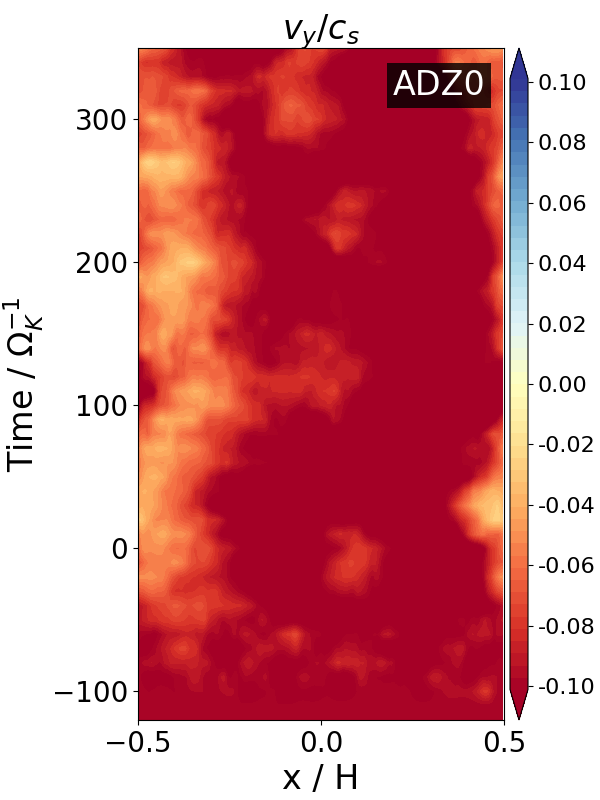}
\includegraphics[width=0.33\textwidth]{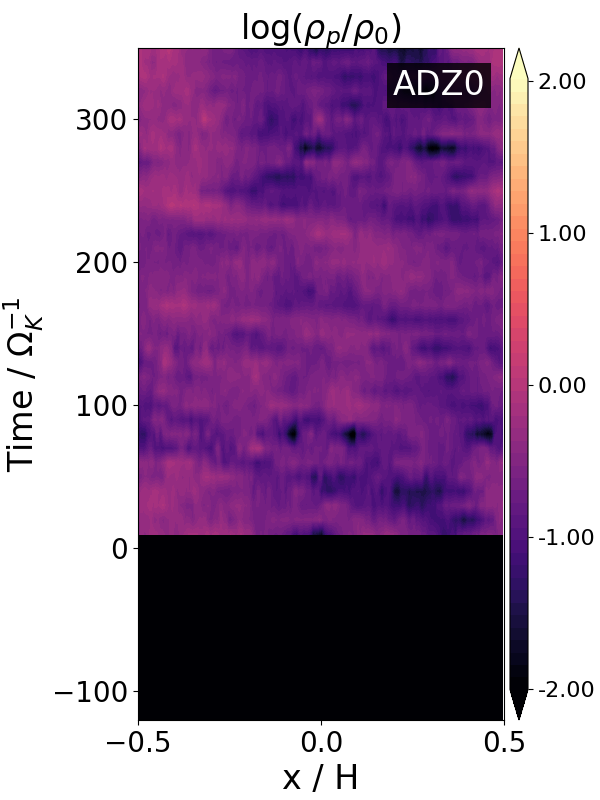}
\includegraphics[width=0.33\textwidth]{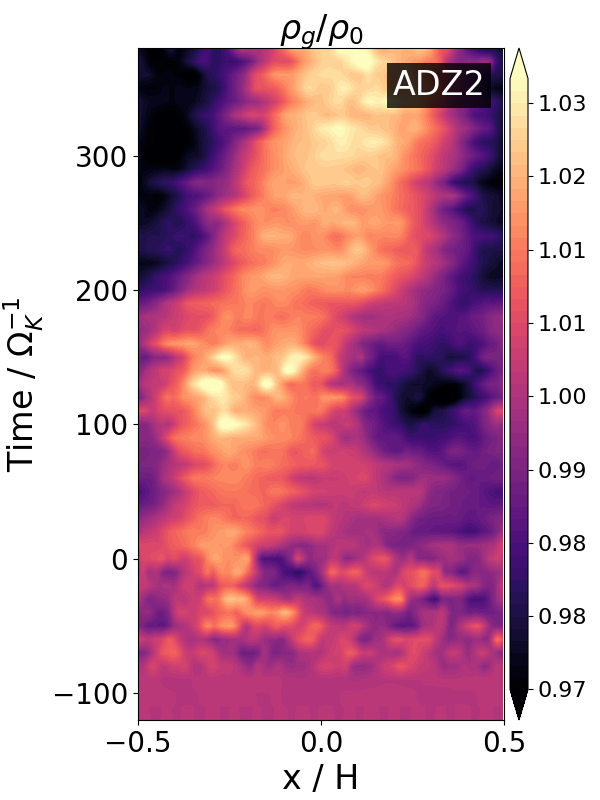}
\includegraphics[width=0.33\textwidth]{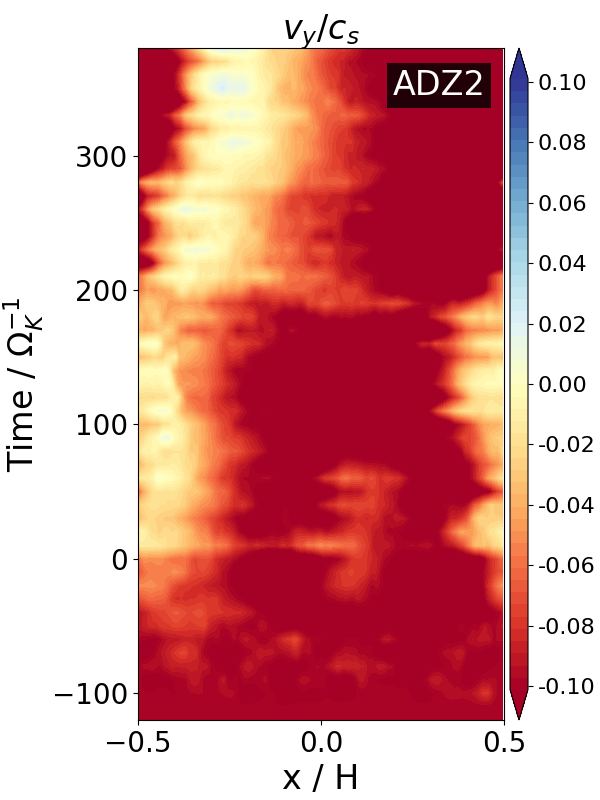}
\includegraphics[width=0.33\textwidth]{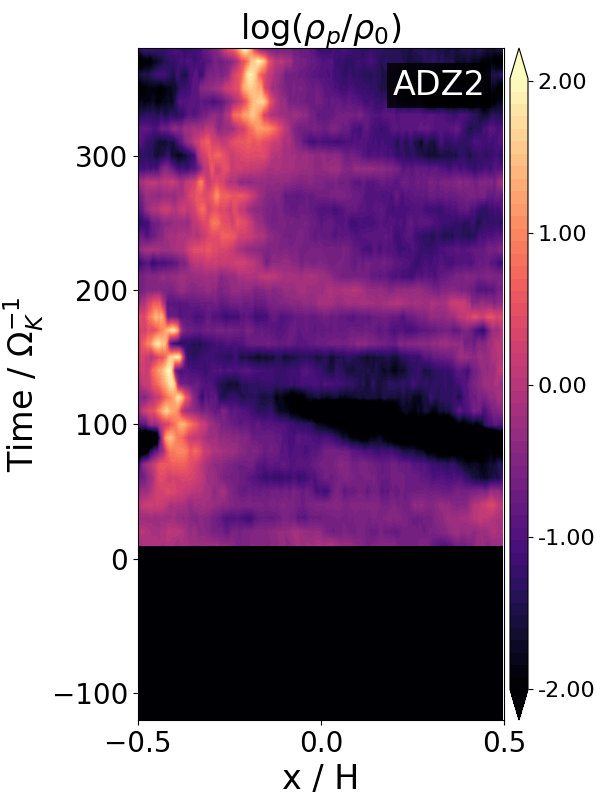}

\caption{\label{fig:x1_time} Time evolution of the azimuthally and vertically averaged radial profiles of gas density (left), Keplerian-subtracted gas azimuthal velocity (middle), and dust density (right) at the midplane ($\pm 1$ cell from $z=0$), for the simulation run ADZ0 (upper panels) and ADZ2 (lower panels). Dust densities for ADZ0 run is scaled to Z=0.02 for easy comparison.
Note that $T=0$ stands for the time of particle injection, and our first particle output is at $T=10\Omega_K^{-1}$.
}
\end{figure*}

\subsection{Clumping vs. pressure variations} \label{subsec:zonalflow}

Apparently, our finding on particle clumping challenges results from recent studies \citep[e.g.][]{umurhan20,chen20,gole20} that turbulence damp the SI and thus suppress dust clumping.
To further diagnose the reason why MRI turbulence promotes clumping, and the clumping process in general,
we present in Figure \ref{fig:x1_time} the time evolution of azimuthally and vertically averaged radial profiles of gas density, gas azimuthal velocity, and dust density for run ADZ0 and ADZ2. The formation and evolution of the dust clumps in ADZ2 are seen in the lower-right panel at $T=80\sim 180 \Omega_K^{-1}$ and $T=330\sim 380 \Omega_K^{-1}$, corresponding to the periods of dust maximum density reaching $\sim 1000$ in Figure \ref{fig:hpdmax_hist}.

One important feature in the MRI turbulence is the spontaneous formation of long-lived (tens of orbits) zonal flows, corresponding to large-scale, quasi-axisymmetric density (and hence pressure) variations \citep{fromang05,johansen06,johansen09a}. These large-scale gas pressure variations can be seen in the left panels of Figure \ref{fig:x1_time}.
The pressure variations change the pressure gradient profile, alter particle radial drift and potentially trap particles at the local pressure maxima. Note that because we implement background pressure gradient by a radial force, the actual local pressure maximum, if present, will appear slightly to the negative-$x$ (left) side of the density maximum.

A more straightforward way to examine the correspondence of particle clumping and pressure variations is by analyzing the gas azimuthal velocity. A positive pressure gradient leads to super-Keplerian rotation, corresponding to $v_y>0$ in our simulation with Keplerian shear subtracted. A negative pressure gradient corresponds to $v_y<0$. Hence, the local pressure maximum is marked by $v_y=0$ regions, with $v_y>0$ on the left side and $v_y<0$ on the right. The gas azimuthal velocity $v_y$ versus time is shown in the middle panels of Figure \ref{fig:x1_time}
\footnote{Our formulation mimics the outward pressure gradient in the gas by an inward radial force on the particles (see \S \ref{subsec:formulation}), making gas azimuthal velocity $\eta v_K=\Pi c_s$ larger than the actual case. In Figure \ref{fig:x1_time}, we have subtracted this component to show $v_y$ in the actual situation.}.

In ADZ0 where no clumping is present, we see that pressure variation from the zonal flow is not strong enough to overcome the background pressure gradient, and gas azimuthal velocity $v_y$ is always negative (indicating sub-Keplerian rotation) throughout the simulation box.
With dust feedback included in ADZ2, on the other hand, pressure variation is enhanced. It successfully overcomes the background gradient to generate a pressure maximum ($v_y=0$ region) to the left of the density peak, and particle clumping is seen in regions with $v_y=0$. Although it is less clear why this is the case, our results indicate that dust feedback can promote clumping by enhancing the gas pressure bump in the zonal flow.
Moreover, we can infer that the disruption of the clump around $T=200\Omega_K^{-1}$ is associated with the weakening of the zonal flow, where a pressure maxima is no longer sustained.

Overall, we have identified that the formation of pressure maxima and dust feedback are the two essential elements for dust clumping, and that dust feedback appears to enhance zonal flows and promote the formation of pressure maxima.
Dust clumping and planetesimal formation thus follow a two-stage process with particle concentration in pressure maxima followed by clumping due dust feedback.

What is yet to be clarified is the interpretation on the role of dust feedback. It is commonly taken into granted that incorporation of dust feedback leads to clumping by the SI. However, dust feedback does not necessarily lead to clumping (solid abundance must exceed some threshold), and on the other hand,
the SI is not supposed to operate at pressure maxima (its free energy comes from the pressure gradient). While small pressure gradient promotes dust clumping for pure SI \citep{bai10c}, we do not observe dust clumping when pressure variation from the zonal flow is close but not yet to overcome background pressure gradient (e.g., around time $T\sim200\Omega_K^{-1}$ in run ADZ2 where $\rho_{\rm max}$ plunges, also see \S \ref{subsec:pgrad} and Figure \ref{fig:x1_time_more}).
Alternatively, one may also argue that our results are in line with the notion that the SI is suppressed by modest-to-strong level of turbulence \citep{gole20}. Yet, the requirement of dust feedback for clumping remains to be addressed.

Similar results were obtained in ideal MHD simulations of \citet{johansen11}, where they attribute the clumping to dust trapping in pressure ``bumps" generated by the zonal flow with the SI being a secondary effect.  Nevertheless, zonal flows in their ideal MHD simulations appear more intermittent with lower amplitudes than ours, and they did not address whether pressure variations in their simulations actually overcome background pressure gradient to become bumps.
Our analysis from more realistic setting with AD-dominated MRI turbulence in the outer PPDs allows us to clearly identify the occurrence of clumping with pressure maxima, and we will present further study in \S\ref{subsec:pgrad} which will strengthen our conclusions.

\subsection{Properties of particle clumps} \label{subsec:parclumping}

\begin{figure}[t]
\includegraphics[width=0.48\textwidth]{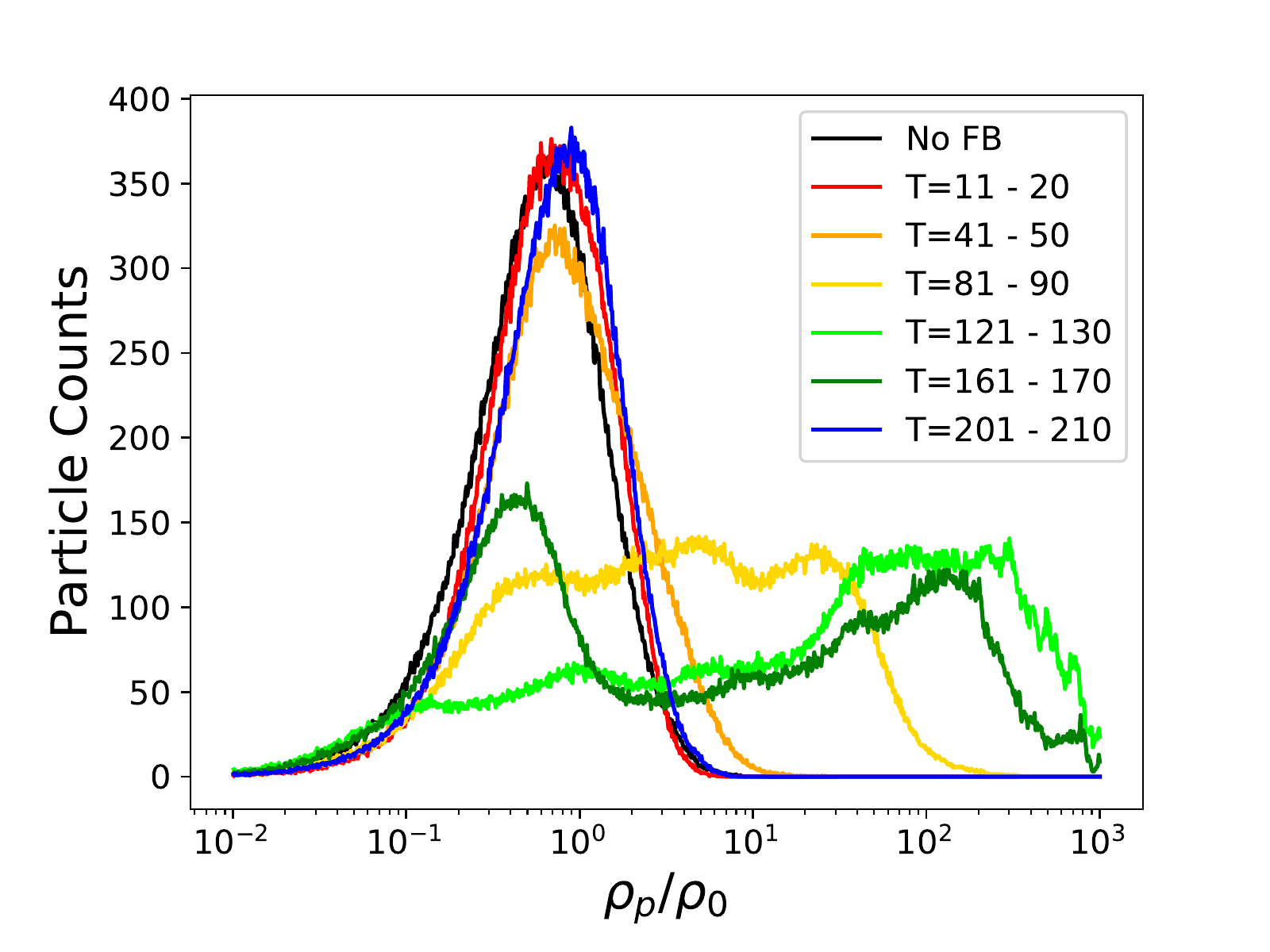}

\caption{\label{fig:cpdf} Distribution function of dust density in runs ADZ0 (no feedback) and ADZ2 ($Z=0.02$). For run ADZ2, the distribution function is measured at different simulation times $T$ in the unit of $\Omega_K^{-1}$, and marked in different colors (see legend). Each distribution function is time-averaged within $10\Omega_K^{-1}$. For run ADZ0, the distribution function is measured and time-averaged between $T=11-20\Omega_K^{-1}$.}
\end{figure}

In the lower panel of Figure \ref{fig:hpdmax_hist}, we identify clumping events by the maximum local dust density.
In order to further analyze the properties in the particle clumps, we show in Figure \ref{fig:cpdf} the distribution function of dust density. The local dust density $\rho_p$ is uniformly divided into 1000 bins in logarithmic space between $0.01\rho_0$ and $1000\rho_0$, and the distribution function is measured by counting the number of dust particles located in a region with local $\rho_p$ in each bin.

We take the first clumping period in ADZ2 ($T=0-210\Omega_K^{-1}$) for analysis, and measure the distribution function every $40 \Omega_K^{-1}$ (except that the first measurement is taken at $T=10\Omega_K^{-1}$ to avoid numerical artifacts at the beginning of inserting particles). Each distribution function is time-averaged within 10$\Omega_K^{-1}$, in order to reasonably reduce the noise level.
For comparison, we also present a distribution function of run ADZ0 measured at $T=11-20\Omega_K^{-1}$, same time period as the first function measured in ADZ2. Due to the absence of dust clumping, the distribution function for ADZ0 is generally consistent throughout simulation time.

The whole process of clump formation and evolution in ADZ2 is well illustrated by the $\rho_p$ distribution function.
The run starts with a smooth Poisson-like distribution prior to clumping ($T=11-20\Omega_K^{-1}$), with width reflecting the density fluctuations in turbulence. The distribution function peaks at $\rho_p\sim \rho_0$, consistent with the expected dust density at midplane that we initially set up in the simulation, $\rho_{p0} \sim Z \rho_0 H/H_{p0}\sim \rho_0$, where $H_{p0}=0.02$ is the initial particle scale height set in the simulation. The distribution maximizes at $\rho_p\sim 10\rho_0$, consistent with the $max(\rho_p)/\rho_0$ value shown in Figure \ref{fig:hpdmax_hist}.

The profile then flattens and shifts towards higher density as the clumping starts ($T=81-90\Omega_K^{-1}$).
During the time of steady, maximum clumping (e.g. $T=121-130\Omega_K^{-1}$), the density distribution peaks at $\rho_p\sim100\rho_0$, and maximizes at $\rho_p\sim1000\rho_0$, again consistent with Figure \ref{fig:hpdmax_hist}.
After the period of clumping, the particle density distribution shifts back after $T = 161-170\Omega_K^{-1}$, with the distribution profile similar to that before clumping.
This profile before and after clumping is also consistent with the density distribution profile in ADZ0, indicating that dust feedback itself has little effect on particle density distribution properties, and the profile/shape of distribution function is mainly controlled by the presence of strong clumping events.

Overall, during the clumping event, the bulk of the dust density distribution shifts towards higher values, deviates from Poisson profile, and peaks at close to the high-density end of the distribution. This further indicates that in addition to forming the densest clumps of $\rho_p \sim 1000 \rho_0$, a large fraction of particles also reside in relatively dense clumping regions ($\sim 100 \rho_0$).

\section{Radial transport of dust particles} \label{sec:radtrans}

\begin{figure*}[t]

\includegraphics[width=\textwidth]{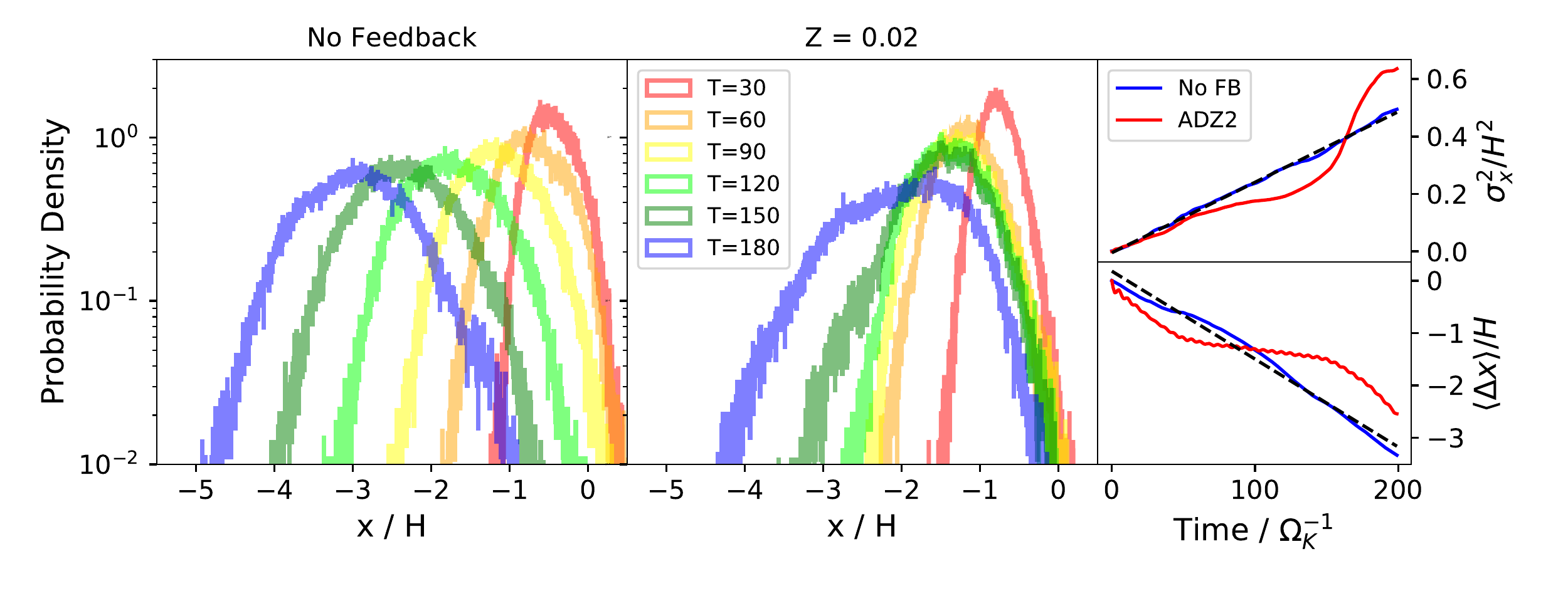}

\caption{\label{fig:radialdiff} Left/middle: probability distribution for particle radial displacements in run ADZ0 (no feedback)/run ADZ2 (Z=0.02). The probability distributions are measured at different simulation times $T$ in the unit of $\Omega_K^{-1}$, and marked in different colors (see legend). Upper right: time evolution of the variance of the particle radial displacement. Lower right: time evolution of the mean particle radial displacement. The black dashed lines in the right panels show the linear fit to the results from run ADZ0.}
\end{figure*}

In this section, we analyze the radial transport of dust particles, including the role of dust clumping on dust transport.
In doing so, we track the radial trajectory $x_i(t)$ of a sample of particles in the simulation.
From the time of particles injection $t_0$, we trace the radial displacement of particles $x_i(t_0+\Delta t)-x_i(t_0)$, and reset particle positions when they cross radial boundaries in order to maintain continuous particle trajectories.

The left and middle panels of Figure \ref{fig:radialdiff} show the distribution of particle radial displacement over time of $T=180\Omega_K^{-1}$ after injecting particles for ADZ0 and ADZ2 runs respectively, corresponding to the first round of particle clumping in run ADZ2 (see the lower panel of Figure \ref{fig:hpdmax_hist}). Without dust feedback, the bulk of particles drift radially inward, and diffuse to exhibit broader radial width over time. In the following subsections, we analyze particle radial drift and diffusion separately.

\subsection{Radial drift} \label{subsec:raddrift}

In laminar disks, for a sub-Keplerian speed $\Delta v_K$, dust particles suffer from radial drift. For particles with dimensionless stopping time of $\tau_s$, and ignore particle backreaction, the radial and azimuthal drift velocity relative to the local Keplerian velocity is given by \citep{nakagawa86}

\begin{equation}
    v_r = -2\frac{\tau_s}{\tau_s^2+1}\Delta v_K\ ,\quad
    v_\phi = -\frac{1}{\tau_s^2+1}\Delta v_K\ .
\end{equation}
For our choice of $\Delta v_K = 0.1 c_s$, dust particles are expected to drift inward with $v_r = -0.02 c_s$.

We measure the mean particle radial displacement $\langle \Delta x\rangle$ as a function of time for ADZ0 and ADZ2 runs, shown in the lower right panel of Figure \ref{fig:radialdiff}.
For ADZ0, the particle displacement increases almost linearly with time, and a linear regression fit estimates the averaged radial drift velocity of $v_r = -0.017 c_s$, very close to the theoretical expectation of $-0.02 c_s$.
We also see that the drift speed also slightly deviate from the linear fit, likely owing to weak zonal flows that slightly alter the radial pressure profile.

For run ADZ2, when dust feedback is considered, the $\langle \Delta x\rangle - T$ curve does not closely follow the theoretical expectation anymore. Around $T \sim 50 - 150 \Omega_K^{-1}$, corresponding to the time period with strongest clumping and particle trapping in zonal flows in the simulation, the radial drift is strongly suppressed. Before and after this period, the slope of the $\langle \Delta x\rangle - T$ curve remains similar to that of ADZ0.

This reduction in dust radial drift in run ADZ2 is also seen in the middle panel of Figure \ref{fig:radialdiff}. Comparing to the smooth shift of particle distributions in run ADZ0, the particle radial distribution in run ADZ2 shows substantial overlap around $T=60-150\Omega_K^{-1}$, indicating the bulk of the particles are trapped and the radial drift is halted.

\subsection{Radial diffusion} \label{subsec:raddiff}

Under the assumption of random walk, the dust radial diffusion coefficient $D_x$ can be measured by

\begin{equation}
    D_x = \frac{1}{2}\frac{d \sigma_x^2}{dt},
\end{equation}
where $\sigma_x$ is the rms of particle radial displacement distribution. We measure the $\sigma_x^2$ as a function of time for runs ADZ0 and ADZ2, presented in the upper right panel of Figure \ref{fig:radialdiff}.

For ADZ0, the $\sigma_x^2 - T$ curve closely follows the linear relation, and by fitting the slope, we obtain $D_x \approx1.2\times10^{-3}$. Assuming the gas diffusion coefficient on the radial direction $D_{g,x}$ is largely the same (given $\tau_s\ll1$), the value is lower than $D_{g,z} = 0.003$ that we measured for ADZ0. This is overall consistent with the result of \citet{zhu15}, where $D_{g,x}$ is generally lower (by a factor of $\sim2$) than $D_{g,z}$ in the MRI turbulence with AD. But we also note with the reduction of turbulent eddy time in run AD2 (\S\ref{subsec:teddy}), the value of $D_{g,z}$ and $D_{g,x}$ becomes comparable.

When dust feedback is included, with the presence of dust clumping, particles
no longer follow normal diffusion. The $\sigma_x^2 - T$ curve thus depends on the starting time of measuring the displacement, and is no longer well characterized by a diffusion coefficient.
In run ADZ2, the $\sigma_x^2 - T$ curve is largely consistent with that of ADZ0 before $T \sim 50 \Omega_K^{-1}$, prior to strong particle clumping.
During the clumping period ($T=50-150 \Omega_K^{-1}$), the $\sigma_x^2 - T$ curve is flatter, mainly due to the bulk of particles trapped in the clump and pressure maxima in the zonal flow, similar to our finding in particle radial drift in \S \ref{subsec:raddrift}.
At the end of clumping, $\sigma_x^2$ is strongly enhanced.
As zonal flow weakens, many particles leak from the left side of the pressure bump, re-enter the box from the right side of the bump, and immediately experience a stronger pressure gradient to gain a faster radial drift velocity than the background, which enlarges $\sigma_x^2$. This can also be seen in the middle panel of Figure \ref{fig:radialdiff}, where particles with largest radial displacement in run ADZ02 start to catch up with those in run ADZ0 at the final time of $T=180\Omega_K^{-1}$. This enhancement of particle drift also contributes to the dust radial drift velocity that is slightly faster than the ADZ0 case before and after dust clumping period of $T \sim 50 - 150 \Omega_K^{-1}$ (see lower right panel of Figure \ref{fig:radialdiff}).

Overall, our analysis of dust radial transport indicates that pressure maxima from zonal flows, even being temporal, can efficiently trap particles and overcome turbulent diffusion.

\section{Discussion}\label{sec:discussion}

\subsection{The effect of background pressure gradient} \label{subsec:pgrad}

\begin{figure}[t]
\includegraphics[width=0.48\textwidth]{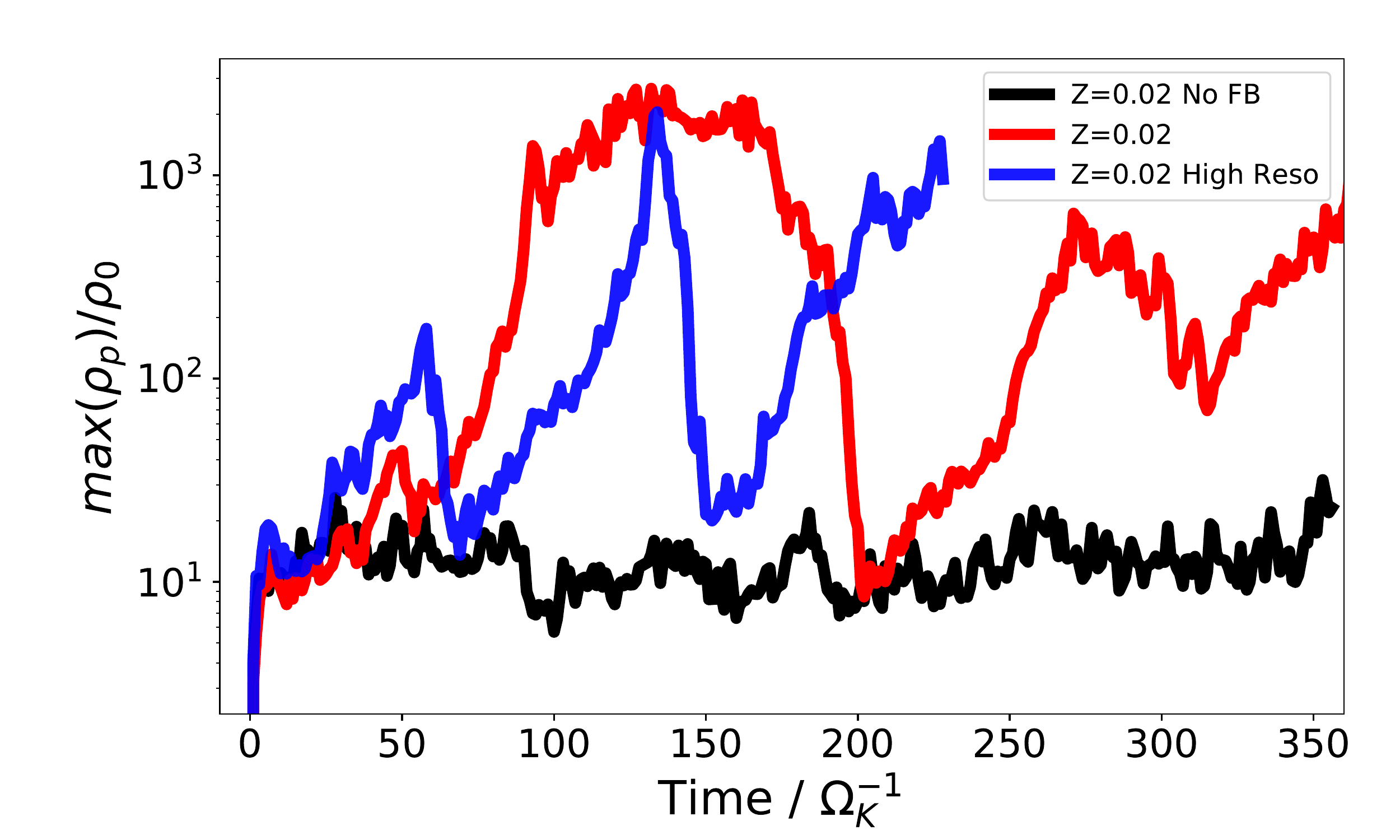}
\includegraphics[width=0.48\textwidth]{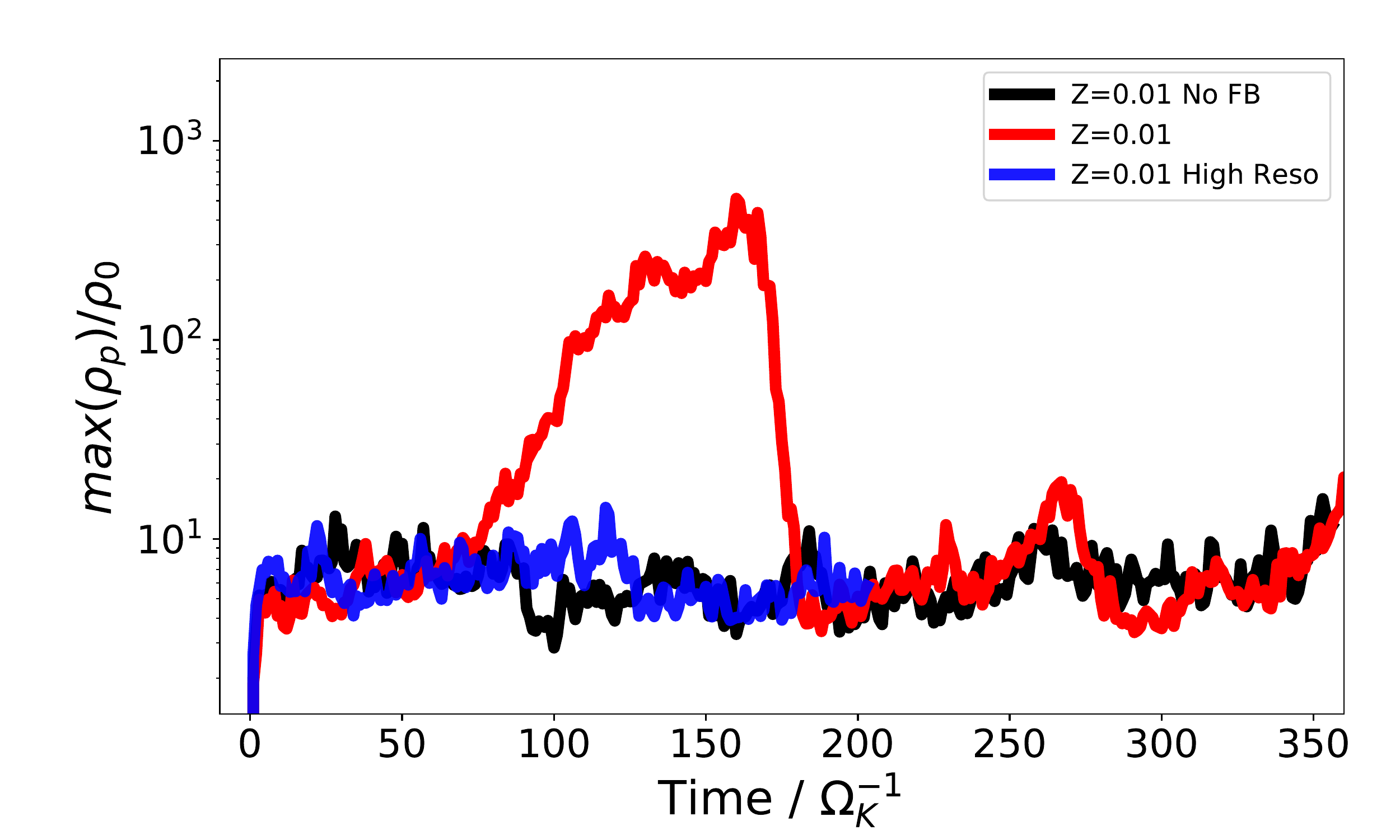}
\includegraphics[width=0.48\textwidth]{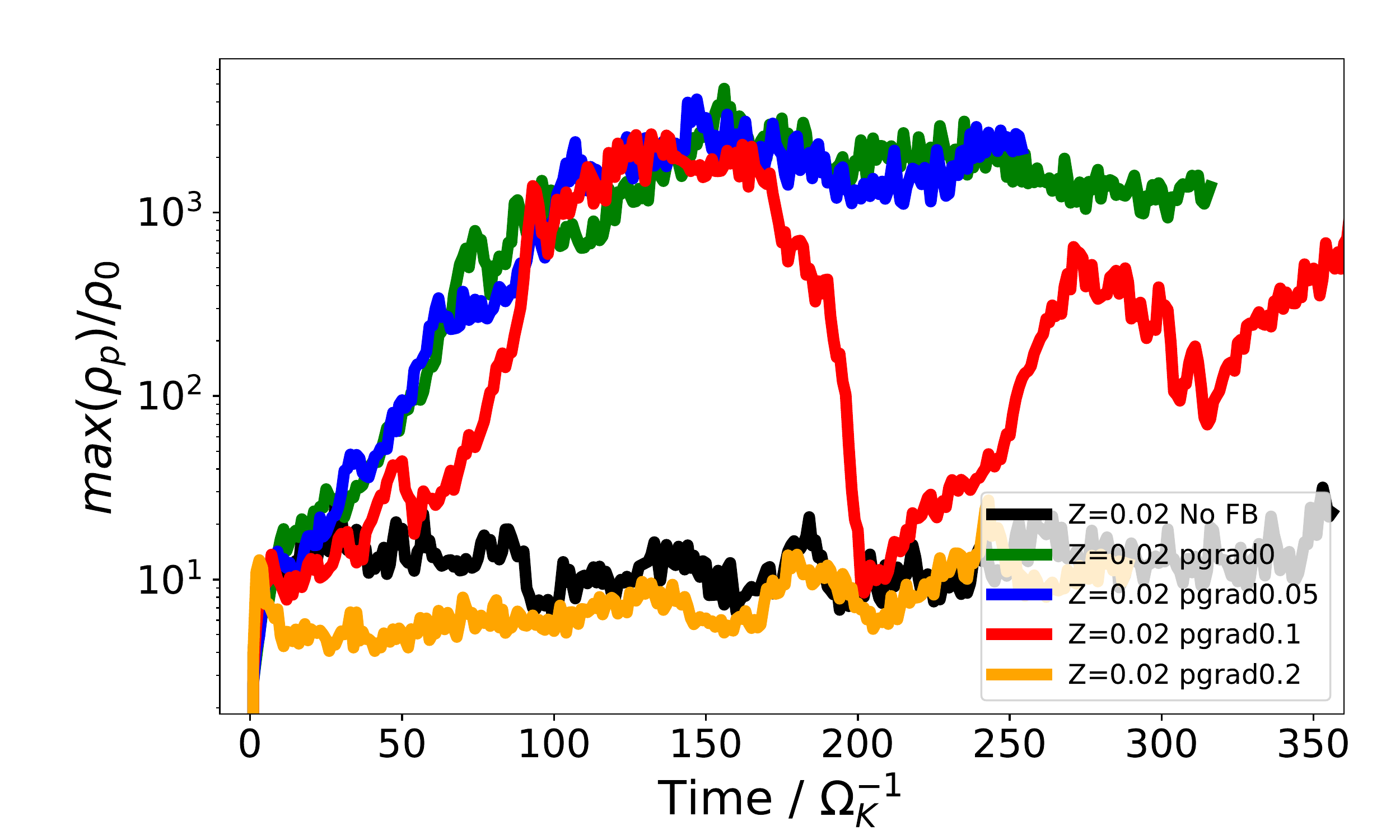}

\caption{\label{fig:morehist} Time evolution of particle maximum density since particle injection, similar to the lower panel of Figure \ref{fig:hpdmax_hist}, but for additional simulation runs. Upper panel: simulations for $Z=0.02$, compared with the higher resolution run (ADZ2H). Middle panel: simulations for $Z=0.01$, compared with higher resolution run (ADZ1H). Lower panel: simulations with different background pressure gradients. ``No FB'' stands for no feedback.
}
\end{figure}

In \S \ref{sec:parclumping}, we show that the particle clumping mechanism in our simulation is different from the conventional SI studies. Particle clumping under the conventional SI scenario is expected to be triggered in the presence of a weak background pressure gradient, while dust clumping under MRI turbulence occurs in the presence of a local pressure maximum, where the pressure variation induced by zonal flow overcomes the background pressure gradient.
In order to further examine this clumping mechanism, we perform runs with different background pressure gradients for comparison.

The lower panel of Figure \ref{fig:morehist} shows the time evolution of the local maximum particle density in simulations with pressure gradients of $\Pi = 0$, $0.05$, $0.1$, and $0.2$, separately.
No particle clumping is seen under $\Pi=0.2$, the strongest background pressure gradient that we test, and clumping is stronger towards lower pressure gradient.
Especially, the strongest particle clumping is seen in our simulation in the absence of a background pressure gradient.
We note that the overall trend is similar to that with the SI: the critical $Z$ required for clumping increases monotonically with the background pressure gradient $\Pi$ \citep{bai10c}.
Nevertheless, the SI does not directly operate in pressure bumps as its free energy arises from background pressure gradient.
Thus, particle clumping seen in our $\Pi = 0$ run further highlights the different clumping mechanism in our case under MRI turbulence.

\begin{figure*}[t]
\includegraphics[width=0.33\textwidth]{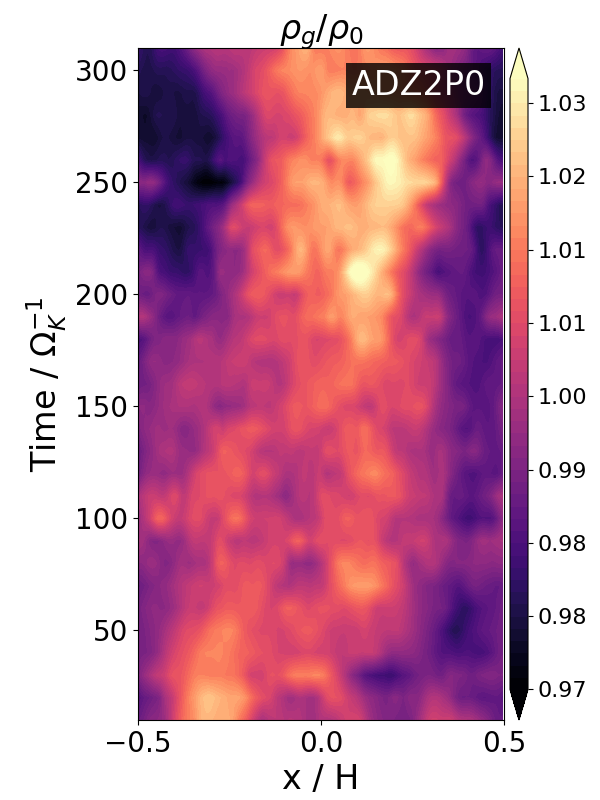}
\includegraphics[width=0.33\textwidth]{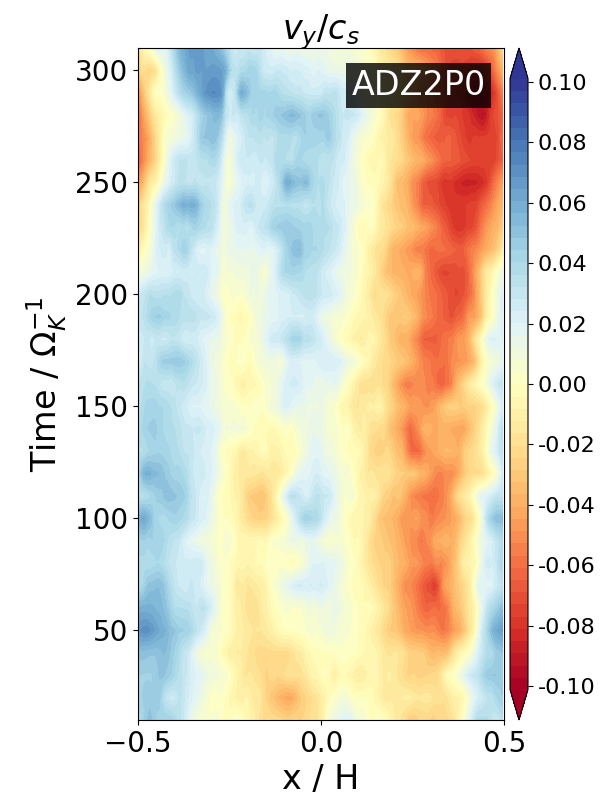}
\includegraphics[width=0.33\textwidth]{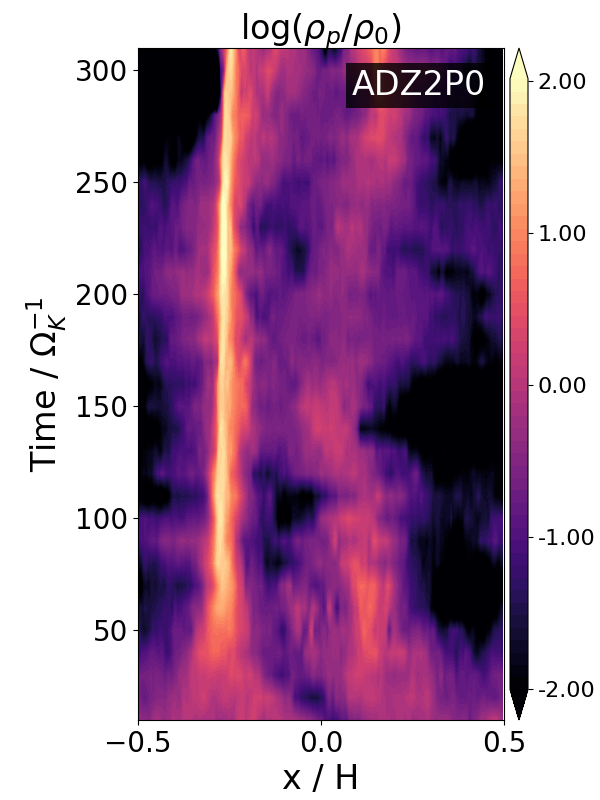}
\includegraphics[width=0.33\textwidth]{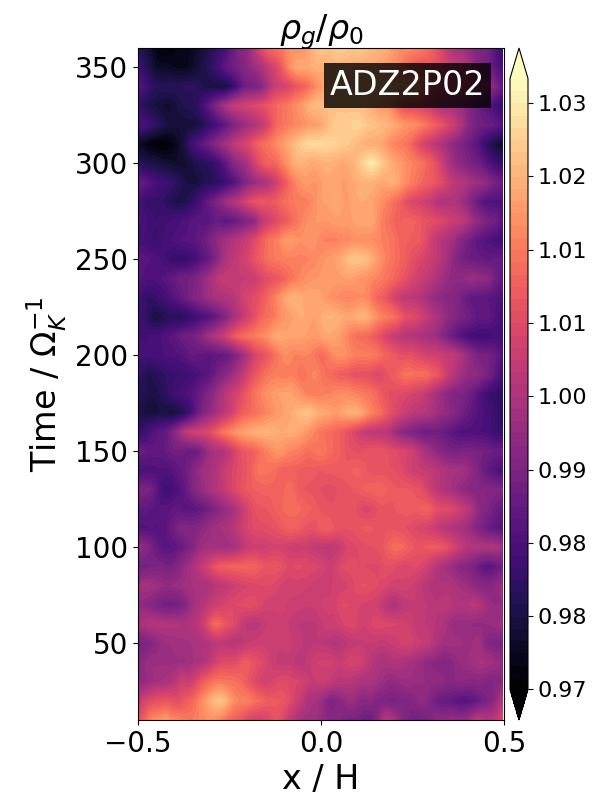}
\includegraphics[width=0.33\textwidth]{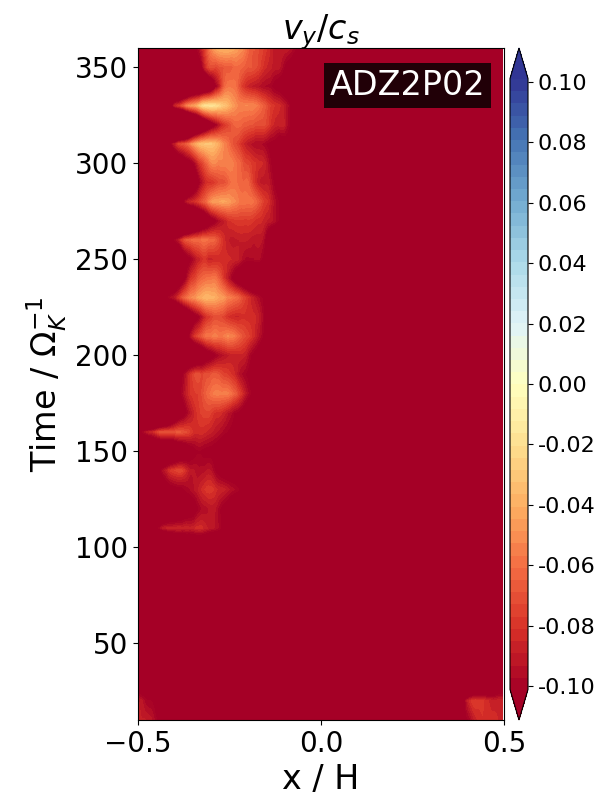}
\includegraphics[width=0.33\textwidth]{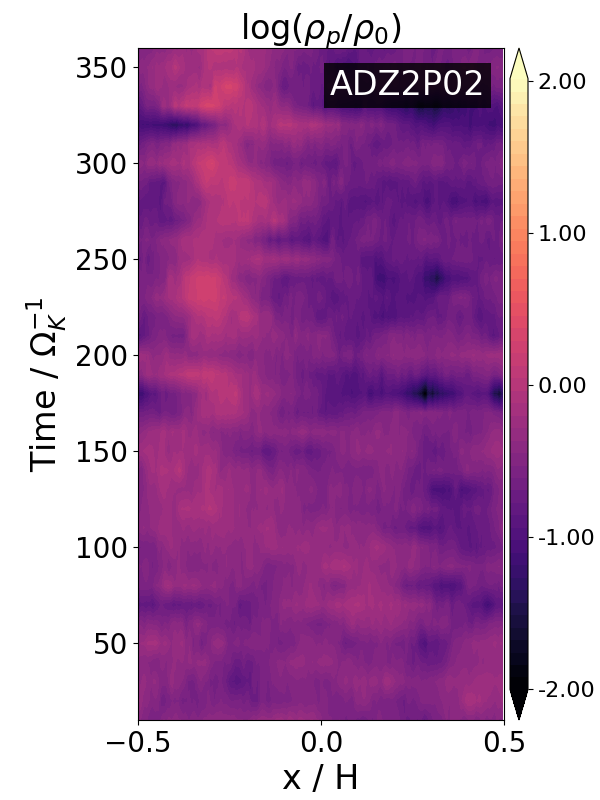}
\includegraphics[width=0.33\textwidth]{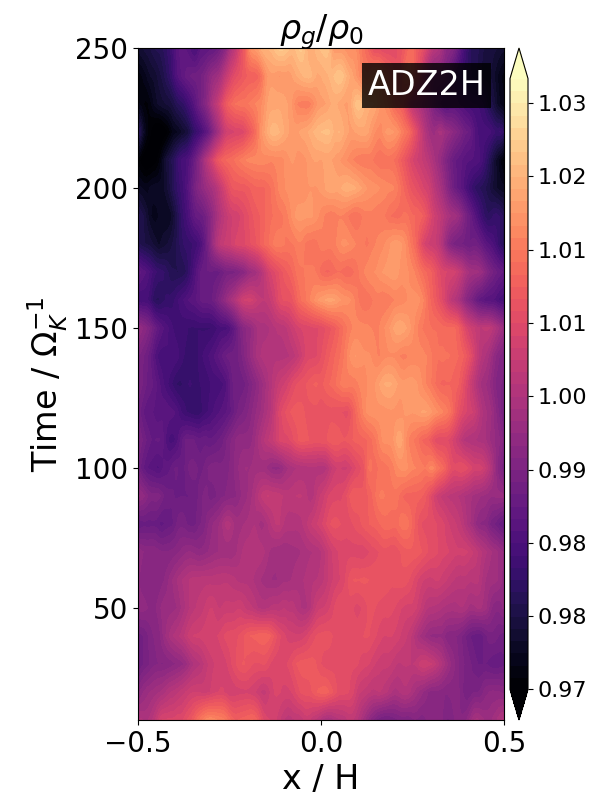}
\includegraphics[width=0.33\textwidth]{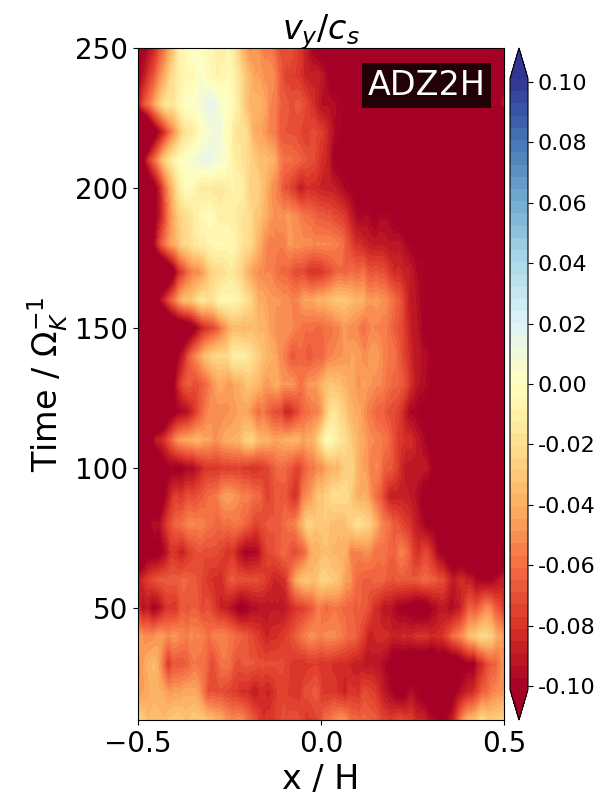}
\includegraphics[width=0.33\textwidth]{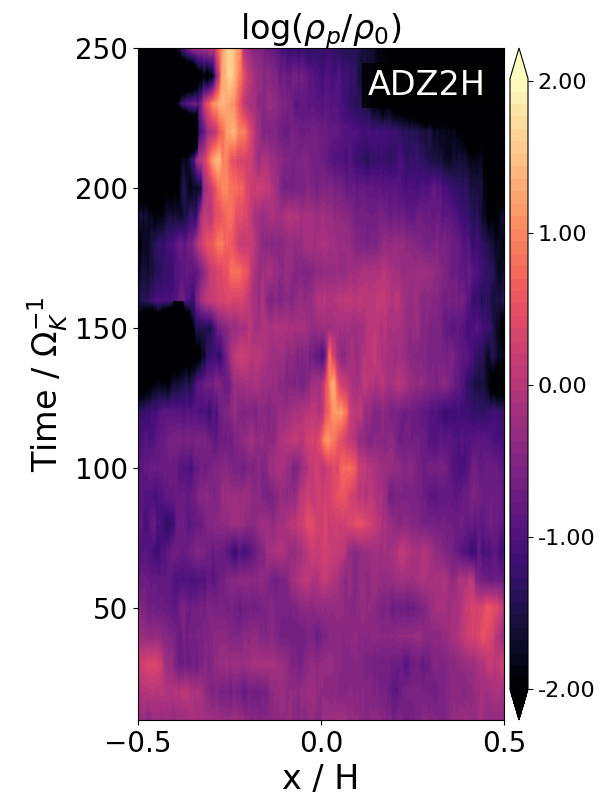}

\caption{\label{fig:x1_time_more} Similar to Figure \ref{fig:x1_time},  but for simulation runs of ADZ2P0 (top), ADZ2P02 (middle), and ADZ2H (bottom).}
\end{figure*}

In \S \ref{subsec:zonalflow}, we illustrated from Figure \ref{fig:x1_time} that the clumping occurs at the local gas pressure maxima, or regions where gas rotation is Keplerian (i.e. $v_y=0$).
Similar to Figure \ref{fig:x1_time}, the upper and middle panels of Figure \ref{fig:x1_time_more} further show the relation between dust clumping and the local pressure maxima, for our runs with different background pressure gradients.

For run ADZ2P0 without background pressure gradient, a pressure variation is formed at $T=0-30\Omega_K^{-1}$ at $x \sim -0.25H$ with an amplitude of $\sim 0.03 \rho_0$, strong enough to generate a pressure maximum (see the corresponding $v_y=0$ region in the middle panel). Although this pressure variation evolves and disappears at later time, it triggers particle clumping, and the clump stays relatively steadily even after the pressure maxima disappeared at later time.
Similarly, a second weak particle concentration is seen at $x \sim 0.1H-0.2H$, with corresponding gas pressure maximum and Keplerian rotation ($v_y=0$) region seen in the simulation box.
A third $v_y=0$ region is seen at $x\sim 0.4H$. However, this region does not correspond to local pressure maximum, but instead a pressure minimum, as the gas rotates at a sub-Keplerian velocity at left side and super-Keplerian at the right. Thus, this $v_y=0$ region does not lead to particle clumping.

In our simulation run with higher pressure gradient $\Pi=0.2$, a pressure bump that is largely similar to that in the ADZ2P0 run is formed, also at $T=0-30\Omega_K^{-1}$, $x \sim -0.25H$, with the same amplitude of $\sim 0.03 \rho_0$. However, no robust particle clumping is seen in the right panel, simply because the pressure bump does not overcome the stronger background pressure gradient, and $v_y$ is overall sub-Keplerian throughout the simulation. The only region with $v_y$ relatively close to Keplerian velocity at later stage of the simulation. It leads to a weak particle concentration, as is seen in a small particle density enhancement shown in the right middle panel of Figure \ref{fig:x1_time_more}.

This exploration of pressure gradients again illustrates that the key reason for dust clumping is the presence of the pressure maxima. The change of background pressure gradient affects particle clumping by altering the overall radial profile of gas pressure and hence the presence of the pressure maxima.

\subsection{The effect of resolution} \label{subsec:reso}

Our fiducial resolution of 256/H fairly well resolves the particle layer ($H_p \sim 0.03H$ for ADZ2) with $\sim$ 16 cells, as well as the width of the particle clumping region ($\lesssim 0.1H$, see e.g. Figure \ref{fig:x1_time}) with $\sim 25$ cells. In our simulations with higher resolution of 384/H, dust clumping properties can be slightly different. The upper panel of Figure \ref{fig:morehist} shows that for Z=0.02, higher resolution may lead to weaker clumping, where particles reach a maximum density similar to the fiducial resolution, but the clumping lasts for a shorter time. For Z=0.01 (middle panel), particle clumping is not seen in the simulation with resolution of 384/H, whereas in fiducial resolution clumping occurs briefly.

The effect of resolution on particle clumping is further shown in the lower panel of Figure \ref{fig:x1_time_more}, the space-time plot for run ADZ2H.
Similar to previously demonstrated (see 5.2 and 7.1), dust clumping is coincident both in time and space with the presence of local pressure maxima ($v_y=0$ regions). Compared to run ADZ2, zonal flow in the high resolution ADZ2H run is overall weaker, and the resulting amplitude of the pressure variation is on average lower.
As a result, the different outcomes in dust clumping we observe in runs with different resolutions likely primarily reflect the non-convergence of zonal flow properties with resolution. We note that in reality, zonal flows properties are tied to disk evolution at global scale, and are not necessarily well characterized in local simulations (see further discussion in \S \ref{subsec:realdisk}). Therefore, we do not pursue further discussion about this non-convergence, but mainly highlight the connection between dust clumping and presence of a pressure maxima.

\subsection{Further collapsing criterion and the effect of turbulence diffusion} \label{subsec:clumpingcrit}

Without external turbulent diffusion, SI simulations with self-gravity show that clumping is followed by gravitational collapse into planetesimals \citep[e.g.,][]{simonj16,abod19,li19}.
Our simulations lack self-gravity. However, with external turbulence, the gravitational collapse of dust clumps may also be suppressed by turbulent diffusion.

\citet{gerbig20} investigated the collapsing criterion considering the competition between self-gravity and turbulent diffusion by both KHI and SI, in addition to the conventional criterion of self-gravity overcoming the stellar tidal shear (i.e., the Roche criterion).
In order to overcome the tidal shear, the dust clump radius $r_{\rm clump}$ must not exceed the Hill length scale, given by their equation (13)

\begin{equation} \label{eq:crithill}
    r_{\rm clump}  < \frac{2 \pi}{9}\frac{G \Sigma_{d,c}}{\Omega_K^2} \equiv L_{\rm Hill}\ ,
\end{equation}
where $\Sigma_{d,c}$ is the surface density of the dust clump.

In order to overcome the turbulent diffusion, the clump radius $r_{\rm clump}$ must be large enough so that the timescale for particles to be diffused over $r_{\rm clump}$ is longer than the gravitational collapsing timescale. Assuming turbulent diffusion coefficient is characterized by $\alpha$, their equation (9) gives

\begin{equation} \label{eq:critdiff}
    r_{\rm clump}  > \frac{\alpha}{\tau_s}\frac{c_s^2}{2\pi G \Sigma_{d,c}} \equiv L_{\rm diff}\ .
\end{equation}

Combining equations (\ref{eq:crithill}) and (\ref{eq:critdiff}) leads to

\begin{equation} \label{eq:critdiff1}
    \frac{3}{2}\sqrt{\frac{\alpha}{\tau_s}} < \frac{\pi G \Sigma_{d,c}}{c_s \Omega_K}\ .
\end{equation}
By adopting Toomre $Q$ parameter \citep{toomre64}

\begin{equation}
    Q \equiv \frac{c_s \Omega_K}{\pi G \Sigma_{g}}\ ,
\end{equation}
and the local metallicity enhancement

\begin{equation}
    \epsilon \equiv \frac{\Sigma_{d,c}}{\Sigma_{d}} = \frac{\Sigma_{d,c}}{Z \Sigma_{g}},
\end{equation}
their equation (70) gives a collapsing criterion of

\begin{equation} \label{eq:qcrit}
    Q < Q_{crit} = \frac{2\epsilon Z}{3}\sqrt{\frac{\tau_s}{\alpha}} = \frac{2\epsilon Z}{3}\frac{H}{H_p}\ .
\end{equation}

In our simulations, the maximum particle density in runs with particle clumping (e.g. ADZ2) is on the order of $10^3$, around 100 times higher than the runs without clumping (e.g. ADZ05, see the lower panel of Figure 2). Adopting this enhancement in particle maximum density of $\epsilon=100$, together with $Z=0.02$, and $H/H_p \sim 20$, analytical prediction by \citet{gerbig20} gives a critical collapsing value of $Q_{crit} \sim 27$, comparable to the Toomre $Q\sim 30$ in MMSN (Weidenschilling 1977, Hayashi 1981).\footnote{The surface density slope in MMSN is steep, with $\Sigma\sim R^{-3/2}$, reaching $\Sigma\approx10$g cm$^{-2}$ at 30AU. More realistic disk surface density profiles are shallower, e.g., \citealp{andrews09}, leading to higher $\Sigma$ at 30 AU and hence smaller $Q$ values.}
Considering the fact that the diffusion $\alpha$ in equation (\ref{eq:qcrit}) is expected to be measured within the dust clump, which could largely deviate from the global  $\alpha$ estimated from $H_p/H$, our estimation here gives an lower limit of $Q_{crit}$.

Under SI-induced turbulence, \citet{klahr20,klahr21} tested the collapsing criterion using simulation with box size comparable to the size of a dust clump, measuring particle diffusivity in the clumping region. They suggested that the level of turbulence within the SI-clumps sets the size of planetesimals. However, as clumping occurs in pressure bumps in our simulations without much free energy to drive the SI, source of turbulence in our (unresolved) clumps is likely dominated by MRI. Though we are not in a position to infer the exact turbulence level at this moment, our study may open up a new avenue for understanding the size distribution of planetesimals.

\subsection{Towards planetesimal formation in realistic disks (and rings)} \label{subsec:realdisk}

In \S \ref{ssec:turb}, we raised the conundrum where external turbulence is disfavored by conventional understanding of planetesimal formation by the SI, yet observations indicate modest level of turbulence by modeling the width of ALMA rings. Our simulations represent the first step towards resolving this conundrum: we have found strong evidence that particle clumping occurs in pressure maxima formed by the MRI zonal flows. To some extent, our results are consistent with the suppression of SI by external turbulence, as we do not find particle clumping occuring in regions with finite pressure gradient. It remains to identify the precise mechanism for particle clumping in pressure maxima, where the SI is not expected to operate, but at least we observe that dust feedback does play a role in enhancing zonal flows and promote clumping (see also analysis by \citealp{auffinger18}).

Our results highlight the importance of investigating planetesimal formation in turbulent pressure bumps. The recent work of \citet{carrera21}, who conducted hydrodynamic simulations of the SI, is towards this direction. However, most of their imposed pressure variations were insufficient to overcome background pressure gradient, and planetesimal formation in their simulations mostly occurs by the SI in regions with low pressure gradient, instead of at the point of pressure maximum (if present). This is no longer the case in our simulations with external MRI turbulence, where clumping occurs just at the pressure maxima. We also note that in the presence of a real pressure bump but without external turbulence, while dust can drive some turbulence and form planetesimals by the SI as they drift towards the pressure maximum, as also found in \citet{carrera21}, such processes are transient and remaining dust particles would eventually settle and concentrate indefinitely towards the midplane of the bump center as free energy from radial drift is exhausted. Such a state is inconsistent with the finite width of ALMA rings. Therefore, we anticipate that in reality, planetesimal formation likely occurs in turbulent pressure bumps. While we do not impose pressure bumps on our own, the MRI zonal flows provides a natural mechanism to create the pressure variations, potentially overcoming the background pressure gradient to form pressure bumps, serving as an initial step towards more realistic understandings.

One major caveat in our simulations is the use of local shearing-box framework without vertical stratification, making the zonal flow properties subject to large uncertainties. Recent global simulations of the MRI with AD confirmed the ubiquity of zonal flows in more realistic disk setting \citep{cui21}. The zonal flows are of modest amplitudes (but higher than ours) with pressure bumps separated by a few disk scale heights, likely compatible with some ALMA ring systems that show finer substructures \citep[e.g.][]{jennings21}.
Therefore, zonal flows from in our simulations are likely not far from more realistic ones, while future investigations of dust dynamics in more realistic pressure bumps are needed to gain further understandings.

\section{Conclusions and Perspectives} \label{sec:conclusions}

In this paper, we conduct high-resolution local shearing-box simulations of hybrid particle-gas MHD to study dust dynamics in weakly turbulent PPDs. We target disk outer regions with turbulence driven by the MRI, incorporating AD as the dominant non-ideal MHD effect, which provides a relatively low but realistic turbulent environment.
Dust backreaction is included, which is believed to be the key ingredient for dust concentration and planetesimal formation in conventional SI theory without external turbulence.
One of the main motivations is to resolve the recently raised conundrum that while the SI is widely believed to be the most promising mechanism for planetesimal formation, it can be damped or suppressed by external turbulence, as recent theoretical and numerical studies have shown \citep{umurhan20,chen20,gole20}. Such turbulence is likely present in PPDs from both observational \citep{dullemond18,rosotti20} and theoretical (e.g., \citealp{simon13,cui21}) perspectives, which fundamentally challenges our understandings of planetesimal formation.

This conundrum is resolved based on our simulation results, which are summarized below:
\begin{itemize}
    \item[1.] Dust feedback promotes dust settling towards the midplane in MRI turbulent disks with AD. This enhancement of settling is due to dust feedback reducing the turbulence correlation time.
    \item[2.] Dust clumping under AD-dominated MRI turbulence is generally easier compared to the conventional SI case under pure hydrodynamics.
    Dust clumping favors higher solid abundance $Z$ in the disk, similar to the case of pure SI, but the threshold $Z$ is lower in AD-dominated MRI turbulence.
    \item[3.] The necessary conditions for dust clumping under MRI turbulence
    include dust feedback and the presence of gas pressure maxima.
    \item[4.] In the AD-dominated MRI turbulence, pressure maxima is achieved by the MRI zonal flow that is strong enough to overcome background pressure gradient. Dust feedback enhances the amplitudes of the zonal flow and promotes clumping.
\end{itemize}

Our results suggest that dust clumping and planetesimal formation is a two-stage process, where dust concentration by pressure maxima followed by clumping due to dust feedback. Nevertheless, how dust feedback leads to clumping is debatable. While it is commonly assumed to be due to the SI, the SI is not expected to operate at the pressure maxima, nor under modest-to-strong level of turbulence.
The exact role of dust feedback, and potentially the SI, on particle clumping in the MRI turbulence, remains open for future investigations.

As a first study towards dust dynamics under more realistic disk environments, our work is subject to several limitations. First, the properties of the zonal flows are not necessarily well captured in local unstratified shearing-box simulations (e.g., \citealp{bs14}). Fortunately, recent full 3D global simulations confirm the existence of such zonal flows that do form pressure bumps \citep{cui21}, with bump separation of a few $H$. While incorporating dust to such global simulations with similar resolution can be extremely computationally costly, this can be mitigated by conducting local simulations and drive a pressure bump by forcing, to be shown in our immediate follow up paper. Second, our simulations have neglected self-gravity and hence cannot directly follow the collapse of dust clumps towards planetesimals. Introducing self-gravity is another natural next step, which would break the scale-free nature of our simulations, and we defer for future studies. Finally, we consider single particle species with fixed stopping time in this work.
Further explorations should include a dust size distribution, where dust of different sizes likely act and feedback differently in turbulent MRI disks with/without pressure bumps. Including a dust size distribution will also allow us to compare and contrast with results of multi-species streaming instability without external turbulence \citep[e.g., ][]{krapp19,schaffer21,zhu21}.

\acknowledgements{
We thank Chao-Chin Yang, Min-Kai Lin, and Guillaume Laibe for constructive conversations and discussions, and Greg Herczeg for overseeing the completion of this work.
We thank the referee, Daniel Carrera, for a detailed and constructive report and helpful discussions.
This project is supported by the National Key R\&D Program of China (No.2019YFA0405100). ZX acknowledges the support of NSFC project 11773002. Numerical simulations are conducted on TianHe-1 (A) at National Supercomputer Center in Tianjin, China,
and on the Orion cluster at Department of Astronomy, Tsinghua University.}
\bibliographystyle{apj}
\bibliography{ms}

\end{document}